\documentclass[letterpaper,english,reprint, aps, prl]{revtex4-1}
\usepackage[T1]{fontenc}
\usepackage[latin9]{inputenc}
\usepackage{babel}
\usepackage{units}
\usepackage{amsmath}
\usepackage{amssymb}
\usepackage{graphicx}
\usepackage[unicode=true,pdfusetitle,
 bookmarks=true,bookmarksnumbered=false,bookmarksopen=false,
 breaklinks=false,pdfborder={0 0 1},backref=false,colorlinks=false]
 {hyperref}

\makeatletter

\providecommand{\tabularnewline}{\\}

\usepackage{xcolor}
\usepackage{graphicx}

\makeatother

\begin{document}
\title{Atomistic $S$-matrix method for numerical simulation of phonon reflection,
transmission and boundary scattering }
\author{Zhun-Yong Ong}
\email{ongzy@ihpc.a-star.edu.sg}

\affiliation{Institute of High Performance Computing, A{*}STAR, Singapore 138632,
Singapore}
\date{\today}
\begin{abstract}
The control of phonon scattering by interfaces is critical to the
manipulation of heat conduction in composite materials and semiconducting
nanostructures. However, one of the factors limiting our understanding
of elastic phonon scattering is the lack of a computationally efficient
approach for describing the phenomenon in a manner that accounts for
the atomistic configuration of the interface and the exact bulk phonon
dispersion. Building on the atomistic Green's function (AGF) technique
for ballistic phonon transport, we formulate an atomistic $S$-matrix
method that treats bulk phonon modes as the scattering channels and
can determine the numerically exact scattering amplitudes for individual
two-phonon processes, enabling a highly detailed analysis of the phonon
transmission and reflection spectrum as well as the directional dependence
of the phonon scattering specularity. Explicit formulas for the individual
phonon reflection, absorption and transmission coefficients are given
in our formulation. This AGF-based $S$-matrix approach is illustrated
through the investigation of: (1) phonon scattering at the junction
between two isotopically different but structurally identical carbon
nanotubes, and (2) phonon boundary scattering at the zigzag and armchair
edges in graphene. In particular, we uncover the role of edge chirality
on phonon scattering specularity and explain why specularity is reduced
for the ideal armchair edge. The application of the method can shed
new light on the relationship between phonon scattering and the atomistic
structure of interfaces.
\end{abstract}
\maketitle

\section{Introduction}

It is well-known that phonon scattering with interfaces and surfaces
modifies phonon trajectories and thermal conductivity at the nanoscale
in insulators and semiconductors,~\citep{DGCahill:APR14_Nanoscale}
and can potentially be exploited to control heat conduction for high
efficiency thermoelectric applications.~\citep{MMaldovan:PRL13_Narrow,GRomano:PRB16_Temperature}
For example, the dramatically lower thermal conductivity in silicon
nanowires has been attributed to the enhanced surface scattering of
phonons~\citep{DLi:APL03_SiNW,PMartin:PRL09_Impact,JLim:NL12_Quantifying}
while molecular dynamics simulations suggest that surface modification
can lower the thermal conductivity of silicon thin films.~\citep{BLDavis:PRL14,SNeogi:ACSNano15_Tuning} 

In spite of the importance of phonon scattering by interfaces and
surfaces for thermal transport, our understanding of the phenomenon~\citep{DLi:NMTE15_Phonon,AMaznev:PRB15_Boundary,KKothari:SciRep17_Phonon}
is constrained by the currently limited range of experimental means
for the direct determination of the spectral transmission coefficients~\citep{CHua:PRB17_Experimental}
and relies heavily on acoustic-based approximations which are valid
only in the long-wavelength limit. For example, the acoustic and diffuse
mismatch theories,~\citep{ETSwartz:RMP89_Thermal} which describe
how incoming phonons are scattered elastically by an interface, are
widely used to estimate the transmission probability of phonons while
variations of Ziman's model of elastic scattering by a rough surface~\citep{JZiman:Book60_Electrons}
are used to model diffuse phonon reflection from boundaries.~\citep{ZAksamija:PRB10_Anisotropy}
However, attempts to simulate elastic phonon scattering atomistically
remain constrained by the lack of an efficient numerical method that
can treat directly the scattering-induced transition between an incoming
bulk phonon and an outgoing bulk phonon of equal frequency on either
side of the interface. Although other atomistic approaches such as
wavepacket-based simulations~\citep{NZuckerman:PRB08_Acoustic} have
been used to study phonon transmission and reflection at interfaces~\citep{PKSchelling:APL02_Phonon}
and surfaces,~\citep{ZWei:JAP12_Wave,CShao:JAP17_Probing} their
application is difficult as they require large simulation domains
and are computational expensive, limiting their usefulness for extracting
quantitative insights as well as applicability to more complicated
atomistic structures. The traditional atomistic Green's function (AGF)
method,~\citep{WZhang:NHT07,JSWang:EPJB08_Quantum,SSadasivam:ARHT14_Atomistic}
which is numerically exact and can be used to compute the overall
transmittance spectrum for solid interfaces,~\citep{WZhang:JHT07_Simulation,ZTian:PRB12_Enhancing}
cannot resolve the transmission of individual phonons.

Nevertheless, there has been significant recent progress~\citep{ZYOng:PRB15_Efficient,SSadasivam:PRB17_Phonon}
in extending the AGF method for studying the transmission and conversion
of \emph{individual} phonon modes at the interface, giving us a more
detailed picture of the forward scattering (or transmission) of phonons
in terms of their polarization and wavelength dependence. Building
on methods developed for tight-binding models of quantum transport
in Ref.~\citep{PAKhomyakov:PRB05_Conductance} and exploiting the
properties of the Bloch matrix,~\citep{TAndo:PRB91_Quantum} it is
shown in Ref.~\citep{ZYOng:PRB15_Efficient} how the transmission
coefficient of individual phonon modes can be calculated by using
an extension of the traditional AGF method. An alternative formulation
of the calculation method that also connects the transmission spectrum
to the bulk phonon spectra and similarly yields the dependence of
the transmission coefficient on phonon polarization and wavelength
is presented in Ref.~\citep{SSadasivam:PRB17_Phonon}. 

Despite their methodological improvements, such AGF-based approaches
remain incomplete because they cannot treat phonon reflection which
is important for understanding the boundary scattering of phonons;
for instance, there is no accessible quantification of phonon polarization
and wavelength conversion in backward scattering (reflection) by the
interface, unlike the case for the forward scattering (transmission)
of phonons. More generally, we lack an \emph{atomistic} approach to
elastic phonon scattering (forward and backward) that considers the
granularity of the crystal structure and can be used for the computation
of scattering cross-sections, important for modeling phonon transport.~\citep{RPrasher:JAP04_Thermal,RPrasher:JAP05_Thermal,NZuckerman:PRB08_Acoustic}
Moreover, in heat conduction within low-dimensional structures such
as atomically thin crystals and nanowires, the specularity of boundary
scattering plays an important role in phonon transport and depends
on the configuration of the boundary.~\citep{PMartin:PRL09_Impact,MBae:NatCommun13_Ballistic,AMajee:PRB16_Length}
Thus, phonon momentum relaxation from elastic boundary scattering
is often invoked~\citep{PMartin:PRL09_Impact} to explain the reduced
thermal conductivity of these nanostructures relative to their bulk
counterparts.~\citep{JLim:NL12_Quantifying} However, this interpretation
relies on certain assumptions about the boundary scattering specularity
and thus, the direct determination of the specularity parameters can
provide a more complete and accurate description of the phenomenon
especially in situations where the atomistic configuration of the
boundary may be important. 

To address this state of affairs, we introduce in this paper a numerical
$S$-matrix approach that generalizes earlier methodological developments
by Ong and Zhang~\citep{ZYOng:PRB15_Efficient} and more importantly,
has the advantage of grounding the description of phonon transmission
and reflection in the language of conventional quantum mechanical
scattering theory, allowing us to draw on existing numerical techniques
and conceptual insights in our modeling of the phenomenon. The key
idea in our method is to exploit the relationship between the Green's
function, which encodes the transition between the initial and final
states~\citep{EEconomou:Book83_Greens} and for which we have well-developed
numerical methods, and scattering theory. To the best of our knowledge,
this conceptual connection between the Green's function and the $S$-matrix
in transport models was first made by Lee and Fisher~\citep{DSFisher:PRB81_Relation}
who describe electron transport through a finite disordered system
in terms of the transmission and reflection of the plane-wave lead
eigenstates. A similar theoretical framework underpins our conceptualization
of phonon transmission and reflection by the interface. Proceeding
along similar lines, we identify the \emph{bulk} phonon modes and
the interface with the scattering channels and scatterer, respectively.
Thus, in the parlance of conventional scattering theory,~\citep{RNewton:Book82_Scattering}
phonon transmission and reflection by the interface is treated as
a multichannel \emph{elastic} scattering problem in which individual
scattering processes are characterized by the scattering amplitudes
between incoming and outgoing phonon channels. 

Nevertheless, although it is known that a formal connection can be
made between the Green's function and scattering, the formulation
of a numerical scheme to determine the elastic scattering amplitudes
between scattering channels remains challenging, because it requires
us to adapt the general scattering formalism, which is largely based
on plane waves,~\citep{EEconomou:Book83_Greens} to variables derived
from the interatomic force constant matrices that characterize the
lattice. In our paper, we give a detailed description of how the scattering
formalism can be implemented numerically in an AGF-based $S$-matrix
approach that uses these interatomic force constant matrices. To minimize
confusion and maintain consistency of notation, the paper is written
in a largely \emph{self-contained} manner so that the basic ideas
behind the calculation techniques are digestible. 

As a cautionary note, we point out that our AGF-based S-matrix method
is only applicable to two-phonon elastic scattering processes. Inelastic
mechanisms such as three-phonon processes,~\citep{PHopkins:JHT11}
which may play a significant role in interfacial thermal transport,
cannot be treated in our approach for now and their treatment requires
complementary approaches like those described in Refs.~\citep{KSaaskilahti:PRB14_Role,YZhou:PRB17_Full}
or possibly a modification of the techniques presented in this paper.
Bearing these limitations in mind, the formalism introduced in this
paper should be sufficiently general that it can be easily extended
to a wider class of problems involving elastic phonon scattering such
as scattering by crystallographic defects (e.g. isotopes, vacancies
and dislocations).

The organization of our paper is as follows. We first review the original
AGF method~\citep{NMingo:PRB03_Phonon} and its extension in Ref.~\citep{ZYOng:PRB15_Efficient}.
Next, we show how the transmitted and reflected phonon modes can be
determined from the incident phonon mode, and derive the transmission
($\bar{\boldsymbol{t}}_{\text{RL}}$ and $\bar{\boldsymbol{t}}_{\text{LR}}$)
and reflection matrices ($\bar{\boldsymbol{r}}_{\text{LL}}$ and $\bar{\boldsymbol{r}}_{\text{RR}}$)
which constitute the \emph{full} $S$ matrix ($\boldsymbol{S}$).
The general properties and application of the transmission and reflection
matrices are also discussed. Finally, the advantages and versatility
of our AGF-based S-matrix approach are illustrated through two examples:
(1) the investigation of phonon reflection and transmission at the
armchair junction between two isotopically different (8,8) carbon
nanotubes, and (2) the investigation of phonon scattering by the graphene
armchair and zigzag edges. In the second example where transverse
periodic boundary conditions are important, we describe the Fourier
decomposition of the force-constant matrices for the efficient computation
of the phonon channels and the application of the zone-unfolding technique~\citep{TBoykin:PRB05_Practical,TBoykin:PhysicaE09_Non}
to map the phonon channels to the phonon modes of the primitive Brillouin
zone of graphene. From our analysis of the effects of edge chirality
and isotopic disorder on phonon specularity, we show why phonon specularity
is reduced for armchair edges. 

\section{Method~\label{sec:Method}}

\subsection{Review of original Atomistic Green's Function (AGF) method}

We briefly give here an overview of the basic elements of the original
atomistic Green's function (AGF) formalism, introduced by Mingo and
Yang in Ref.~\citep{NMingo:PRB03_Phonon}, and its extension developed
in Ref.~\citep{ZYOng:PRB15_Efficient} so that the context for the
new S-matrix method is clear. A similar review of the method can also
be found in Ref.~\citep{ZYOng:JAP18_Tutorial}. The main idea of
the traditional AGF method is as follows: We take the harmonic matrix
$\mathbf{H}$ of the infinite system (left lead, scattering region
and right lead) and break it up into submatrices associated with the
principal layers of the leads and the scattering region. From these
submatrices, we construct: (1) the uncoupled surface Green's function
of each lead and (2) the effective frequency-dependent harmonic matrix
$\mathbf{H}^{\prime}$ of the finite projected system that consists
of the scattering region and its edges. The retarded Green's function
$\boldsymbol{G}^{\text{ret}}$ of the projected system, which determines
overall phonon transmission $\Xi(\omega)$, is then computed from
$\mathbf{H}^{\prime}$. In the extension of the AGF method,~\citep{ZYOng:PRB15_Efficient}
the Bloch matrices and \emph{bulk} phonon modes can be computed from
the uncoupled surface Green's function of the leads and thus used
to determine from $\boldsymbol{G}^{\text{ret}}$ the scattering amplitudes
between the incident and the transmitted modes. 

\subsubsection{Numerical setup for AGF calculation}

\begin{figure}
\newcommand{\SyHC}{$\boldsymbol{H}_{\text{C}}$}
\newcommand{\SyHCL}{$\boldsymbol{H}_{\text{CL}}$}
\newcommand{\SyHCR}{$\boldsymbol{H}_{\text{CR}}$}
\newcommand{\SyHLC}{$\boldsymbol{H}_{\text{LC}}$}
\newcommand{\SyHRC}{$\boldsymbol{H}_{\text{RC}}$}
\newcommand{\SyHLOO}{$\boldsymbol{H}_\text{L}^{00}$}
\newcommand{\SyHROO}{$\boldsymbol{H}_\text{R}^{00}$}
\newcommand{\SyHLOL}{$\boldsymbol{H}_\text{L}^{01}$}
\newcommand{\SyHROL}{$\boldsymbol{H}_\text{R}^{01}$}
\newcommand{\SyHLOR}{$\boldsymbol{H}_\text{L}^{10}$}
\newcommand{\SyHROR}{$\boldsymbol{H}_\text{R}^{10}$}

\includegraphics[scale=0.5]{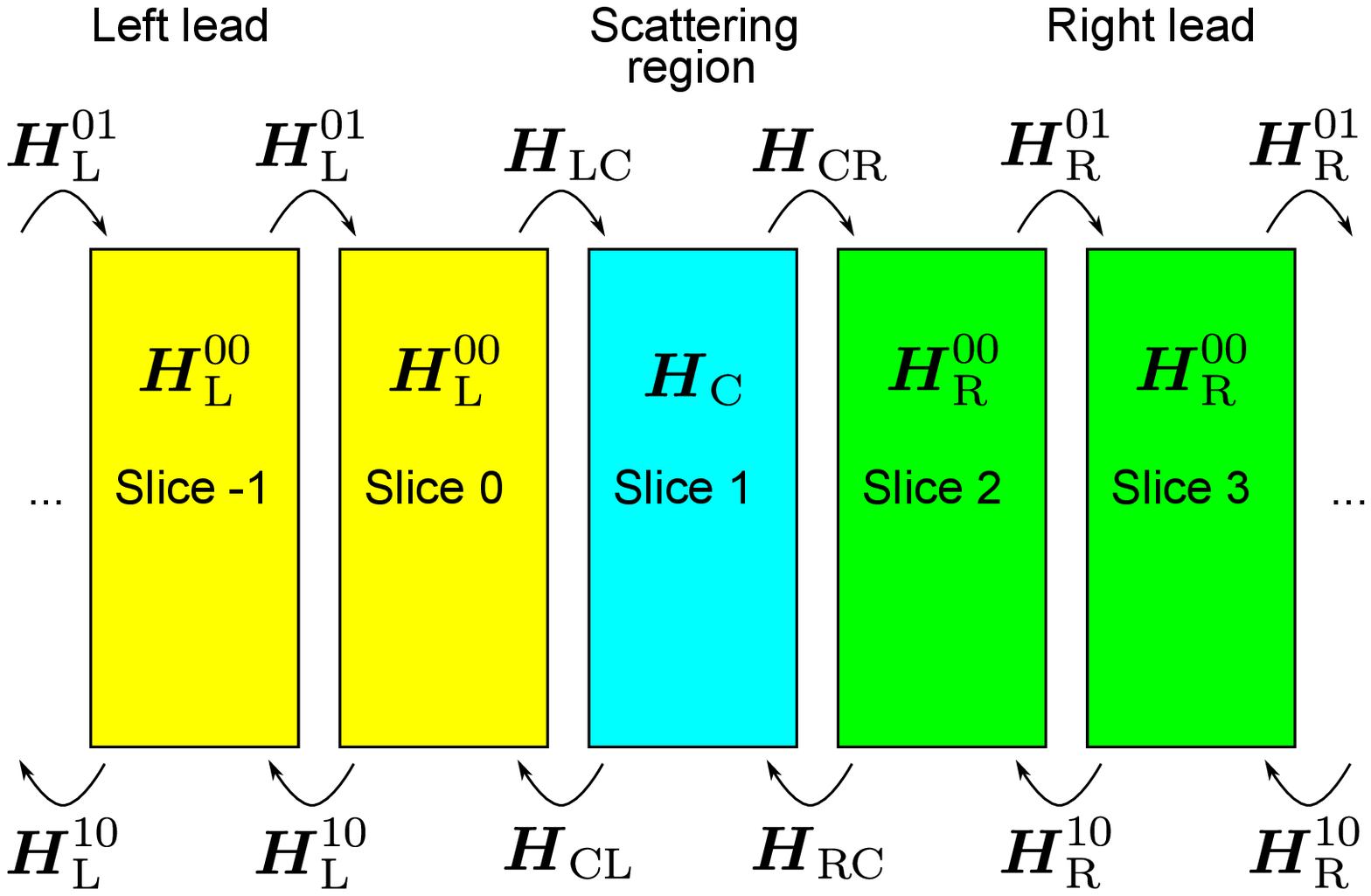}

\caption{Schematic of the scattering system (left lead, scattering region and
right lead) and the submatrices associated with each slice or principal
layer which represents the set of atomic degrees of freedom for a
block row in Eq.~(\ref{eq:SystemForceConstantMatrix}). The left
and right lead each consist of a semi-infinite one-dimensional array
of identical slices while the scattering region corresponds to the
interface.}

\label{fig:SystemSchematic}
\end{figure}
In our scheme, as shown in Fig.~\ref{fig:SystemSchematic}, the system
is partitioned into three subsystems: (1) the left lead, (2) the scattering
region and (3) the right lead. The leads are identified with the physical
bulk lattices while the scattering region contains the interface.
Each lead consists of a semi-infinite one-dimensional array of identical
slices (or principal layers) while the scattering region is considered
a slice by itself. Hence, the entire system has an infinite number
of slices, each of which can be indexed by an integer. The index convention
used in this paper is one in which the index increases as one goes
from left to right. We define the scattering region as slice $1$
while the principal layers in the left and right lead are enumerated
$-\infty,\ldots,0$ and $2,\ldots,+\infty$, respectively. 

Formally, the lattice dynamical properties of the system are determined
by the mass-normalized force-constant matrix $\mathbf{H}$ which represents
the harmonic coupling of the entire system and has the block-tridiagonal
structure, 
\begin{equation}
\mathbf{H}=\left(\begin{array}{ccccccc}
\ddots & \ddots\\
\ddots & \boldsymbol{H}_{\text{L}}^{00} & \boldsymbol{H}_{\text{L}}^{01}\\
 & \boldsymbol{H}_{\text{L}}^{10} & \boldsymbol{H}_{\text{L}}^{00} & \boldsymbol{H}_{\text{LC}}\\
 &  & \boldsymbol{H}_{\text{CL}} & \boldsymbol{H}_{\text{C}} & \boldsymbol{H}_{\text{CR}}\\
 &  &  & \boldsymbol{H}_{\text{RC}} & \boldsymbol{H}_{\text{R}}^{00} & \boldsymbol{H}_{\text{R}}^{01}\\
 &  &  &  & \boldsymbol{H}_{\text{R}}^{10} & \boldsymbol{H}_{\text{R}}^{00} & \ddots\\
 &  &  &  &  & \ddots & \ddots
\end{array}\right)\label{eq:SystemForceConstantMatrix}
\end{equation}
where $\boldsymbol{H}_{\text{C}}$, and $\boldsymbol{H}_{\text{CL}}$
($\boldsymbol{H}_{\text{CR}}$) are respectively the force-constant
submatrices corresponding to the interface region and the coupling
between the interface region and the semi-infinite left (right) lead.
We can associate each slice in Fig.~\ref{fig:SystemSchematic} with
a block row in $\mathbf{H}$. In the standard AGF setup, the block
row submatrices $\boldsymbol{H}_{\alpha}^{00}$ and $\boldsymbol{H}_{\alpha}^{01}$,
where $\alpha=\text{L}$ and $\alpha=\text{R}$ for the left and right
lead, respectively, characterize the lead phonons. If we set the slices
to be large enough so that only adjacent slices can couple, then $\boldsymbol{H}_{\alpha}^{00}$
corresponds to the force-constant matrix for each slice while $\boldsymbol{H}_{\alpha}^{01}$
($\boldsymbol{H}_{\alpha}^{10}$) corresponds to the harmonic coupling
between each slice and the slice to its right (left) in the lead.
In the rest of the paper, we reserve $\alpha$ as the dummy variable
for distinguishing the leads, with $\alpha=\text{L}$ and $\alpha=\text{R}$
representing the left and right lead, respectively. 

We note here that in spite of the infinite number of slices making
up the system, only a finite set of unique force-constant matrices
($\boldsymbol{H}_{\text{C}}$, $\boldsymbol{H}_{\text{CL}}$, $\boldsymbol{H}_{\text{CR}}$,
$\boldsymbol{H}_{\text{L}}^{00}$, $\boldsymbol{H}_{\text{L}}^{01}$,
$\boldsymbol{H}_{\text{R}}^{00}$ and $\boldsymbol{H}_{\text{R}}^{01}$)
are needed as inputs for the AGF calculation because the leads are
made up of identical slices and the Hermiticity of $\mathbf{H}$ implies
that $\boldsymbol{H}_{\text{LC}}=(\boldsymbol{H}_{\text{CL}})^{\dagger}$
and $\boldsymbol{H}_{\text{RC}}=(\boldsymbol{H}_{\text{CR}})^{\dagger}$,
and $\boldsymbol{H}_{\alpha}^{01}=(\boldsymbol{H}_{\alpha}^{10})^{\dagger}$.
The periodic arraying of the slices in the leads means that each slice
constitutes a unit cell and that the bulk phonon dispersion for the
lead can be determined from the expression 
\begin{equation}
\text{det}[\omega^{2}\boldsymbol{I}_{\alpha}-\boldsymbol{D}_{\alpha}(k)]=0\ ,\label{eq:BulkPhononDispersion}
\end{equation}
where $\boldsymbol{D}_{\alpha}(k)=\boldsymbol{H}_{\alpha}^{10}e^{-ika_{\alpha}}+\boldsymbol{H}_{\alpha}^{00}+\boldsymbol{H}_{\alpha}^{01}e^{ika_{\alpha}}$
is the dynamical matrix and $\boldsymbol{I}_{\alpha}$ is the identity
matrix of the size as $\boldsymbol{H}_{\alpha}^{00}$; the variables
$k$ and $a_{\alpha}$ represent the phonon wave vector and lattice
constant in one dimension, respectively. 

\subsubsection{Total phonon transmission}

\begin{figure}

\newcommand{\SyHC}{$\boldsymbol{H}_{\text{C}}^\prime$}
\newcommand{\SyHL}{$\boldsymbol{H}_{\text{L}}^\prime$}
\newcommand{\SyHR}{$\boldsymbol{H}_{\text{R}}^\prime$}
\newcommand{\SyHCL}{$\boldsymbol{H}_{\text{CL}}^\prime$}
\newcommand{\SyHCR}{$\boldsymbol{H}_{\text{CR}}^\prime$}
\newcommand{\SyHLC}{$\boldsymbol{H}_{\text{LC}}^\prime$}
\newcommand{\SyHRC}{$\boldsymbol{H}_{\text{RC}}^\prime$}

\newcommand{\SyUadvL}{$\boldsymbol{U}_\text{L}^\text{adv}(-)$}
\newcommand{\SyUretL}{$\boldsymbol{U}_\text{L}^\text{ret}(-)$}
\newcommand{\SyUadvR}{$\boldsymbol{U}_\text{R}^\text{adv}(+)$}
\newcommand{\SyUretR}{$\boldsymbol{U}_\text{R}^\text{ret}(+)$}

\newcommand{\SyVadvL}{$\boldsymbol{V}_\text{L}^\text{adv}(-)$}
\newcommand{\SyVretL}{$\boldsymbol{V}_\text{L}^\text{ret}(-)$}
\newcommand{\SyVadvR}{$\boldsymbol{V}_\text{R}^\text{adv}(+)$}
\newcommand{\SyVretR}{$\boldsymbol{V}_\text{R}^\text{ret}(+)$}

\includegraphics[scale=0.5]{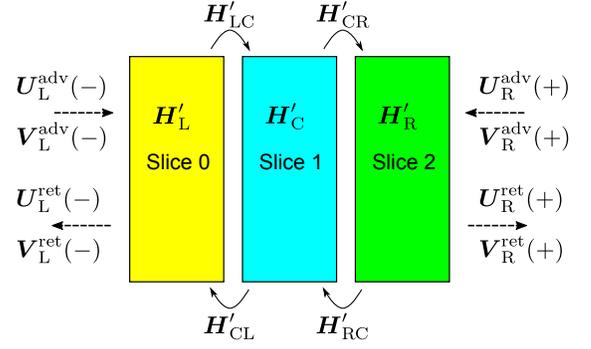}

\caption{Schematic of the finite projected system in Eq.~(\ref{eq:ProjectedForceConstantMatrix}),
consisting of the scattering region (slice $1$) and its terminated
edges (slices $0$ and $2$). The frequency-dependent dynamics of
the semi-infinite leads are implicitly included in $\boldsymbol{H}_{\text{L}}^{\prime}$
and $\boldsymbol{H}_{\text{R}}^{\prime}$ through the surface Green's
functions $\boldsymbol{g}_{\text{L},-}^{\text{ret}}$ and $\boldsymbol{g}_{\text{R},+}^{\text{ret}}$
from which we can derive the incoming and outgoing phonon modes {[}$\boldsymbol{U}_{\text{L}}^{\text{adv/ret}}(-)$
and $\boldsymbol{U}_{\text{R}}^{\text{adv/ret}}(+)${]} and their
group velocities {[}$\boldsymbol{V}_{\text{L}}^{\text{adv/ret}}(-)$
and $\boldsymbol{V}_{\text{R}}^{\text{adv/ret}}(+)${]}.}

\label{fig:ProjectedSystemSchematic}
\end{figure}

In principle, the system dynamics are determined by the infinitely
large $\mathbf{H}$ in Eq.~(\ref{eq:SystemForceConstantMatrix}).
However, for the \emph{effective} dynamics at a fixed frequency $\omega$,
the lattice dynamics problem becomes more tractable as we need only
to project the dynamics onto a finite portion of the system,~\citep{JSWang:EPJB08_Quantum,NMingo:Springer09}
corresponding to slices 0 to 2 in Fig.~\ref{fig:SystemSchematic},
to determine phonon transmission through the scattering region (slice
1). Hence, we use the submatrices in Eq.~(\ref{eq:SystemForceConstantMatrix})
to construct the \emph{effective} harmonic matrix for this subsystem~\citep{JSWang:EPJB08_Quantum}
\begin{equation}
\mathbf{H}^{\prime}=\left(\begin{array}{ccc}
\boldsymbol{H}_{\text{L}}^{\prime} & \boldsymbol{H}_{\text{LC}}^{\prime} & 0\\
\boldsymbol{H}_{\text{CL}}^{\prime} & \boldsymbol{H}_{\text{C}}^{\prime} & \boldsymbol{H}_{\text{CR}}^{\prime}\\
0 & \boldsymbol{H}_{\text{RC}}^{\prime} & \boldsymbol{H}_{\text{R}}^{\prime}
\end{array}\right)\ ,\label{eq:ProjectedForceConstantMatrix}
\end{equation}
where $\boldsymbol{H}_{\text{L}}^{\prime}=\boldsymbol{H}_{\text{L}}^{00}+\boldsymbol{H}_{\text{L}}^{10}\boldsymbol{g}_{\text{L},-}^{\text{ret}}\boldsymbol{H}_{\text{L}}^{01}$
and $\boldsymbol{H}_{\text{R}}^{\prime}=\boldsymbol{H}_{\text{R}}^{00}+\boldsymbol{H}_{\text{R}}^{01}\boldsymbol{g}_{\text{R},+}^{\text{ret}}\boldsymbol{H}_{\text{R}}^{10}$
represent the left and right edge, respectively while $\boldsymbol{H}_{\text{C}}^{\prime}=\boldsymbol{H}_{\text{C}}$
and $\boldsymbol{H}_{\text{CL/CR}}^{\prime}=\boldsymbol{H}_{\text{CL/CR}}=(\boldsymbol{H}_{\text{LC/RC}}^{\prime})^{\dagger}$
(see Fig.~\ref{fig:ProjectedSystemSchematic}). The retarded surface
Green's functions $\boldsymbol{g}_{\text{L},-}^{\text{ret}}$ and
$\boldsymbol{g}_{\text{R},+}^{\text{ret}}$ are given by \begin{subequations}
\\
\begin{equation}
\boldsymbol{g}_{\alpha,-}^{\text{ret}}=[(\omega^{2}+i\eta)\boldsymbol{I}_{\alpha}-\boldsymbol{H}_{\alpha}^{00}-\boldsymbol{H}_{\alpha}^{10}\boldsymbol{g}_{\alpha,-}^{\text{ret}}\boldsymbol{H}_{\alpha}^{01}]^{-1}\label{eq:RetardedLeftSurfaceGF}
\end{equation}
\begin{equation}
\boldsymbol{g}_{\alpha,+}^{\text{ret}}=[(\omega^{2}+i\eta)\boldsymbol{I}_{\alpha}-\boldsymbol{H}_{\alpha}^{00}-\boldsymbol{H}_{\alpha}^{01}\boldsymbol{g}_{\alpha,+}^{\text{ret}}\boldsymbol{H}_{\alpha}^{10}]^{-1}\label{eq:RetardedRightSurfaceGF}
\end{equation}
\label{eq:AllRetardedSurfaceGF}\end{subequations} where $\eta$
is the small infinitesimal part that we add to $\omega^{2}$ to impose
causality, and they are commonly generated using the decimation technique~\citep{FGuinea:PRB83_Effective}
or by solving the generalized eigenvalue equation.~\citep{JSWang:EPJB08_Quantum,SSadasivam:PRB17_Phonon}
Physically, Eq.~(\ref{eq:RetardedLeftSurfaceGF}) is the retarded
surface Green's function for a decoupled semi-infinite lattice extending
infinitely to the left (denoted by the `-' in the subscript of $\boldsymbol{g}_{\alpha,-}^{\text{ret}}$)
while Eq.~(\ref{eq:RetardedRightSurfaceGF}) is the corresponding
surface Green's function for a decoupled semi-infinite lattice extending
infinitely to the right (denoted by the `+' in the subscript of $\boldsymbol{g}_{\alpha,+}^{\text{ret}}$).
In addition, the advanced surface Green's functions can be obtained
from the Hermitian conjugates of Eq.~(\ref{eq:AllRetardedSurfaceGF}),
i.e. $\boldsymbol{g}_{\alpha,-}^{\text{adv}}=(\boldsymbol{g}_{\alpha,-}^{\text{ret}}){}^{\dagger}$
and $\boldsymbol{g}_{\alpha,+}^{\text{adv}}=(\boldsymbol{g}_{\alpha,+}^{\text{ret}})^{\dagger}$. 

To find the phonon transmission through the interface, we compute
the corresponding Green's function for Eq.~(\ref{eq:ProjectedForceConstantMatrix}),
$\boldsymbol{G}^{\text{ret}}=[(\omega^{2}+i\eta)\mathbf{I}^{\prime}-\mathbf{H}^{\prime}]^{-1}$
where $\mathbf{I}^{\prime}$ is an identity matrix of the same size
as $\mathbf{H}^{\prime}$; the $\boldsymbol{G}^{\text{ret}}$ matrix
can be partitioned into submatrices in the same manner as $\mathbf{H}^{\prime}$,
i.e.
\begin{equation}
\boldsymbol{G}^{\text{ret}}=\left(\begin{array}{ccc}
\boldsymbol{G}_{\text{L}}^{\text{ret}} & \boldsymbol{G}_{\text{LC}}^{\text{ret}} & \boldsymbol{G}_{\text{LR}}^{\text{ret}}\\
\boldsymbol{G}_{\text{CL}}^{\text{ret}} & \boldsymbol{G}_{\text{C}}^{\text{ret}} & \boldsymbol{G}_{\text{CR}}^{\text{ret}}\\
\boldsymbol{G}_{\text{RL}}^{\text{ret}} & \boldsymbol{G}_{\text{RC}}^{\text{ret}} & \boldsymbol{G}_{\text{R}}^{\text{ret}}
\end{array}\right)\ .\label{eq:FiniteGreensFunction}
\end{equation}
In the original AGF method,\citep{WZhang:NHT07,JSWang:EPJB08_Quantum}
the phonon transmittance through the scattering region is given by
the well-known Caroli formula:~\citep{CCaroli:JPhysC71,WZhang:NHT07,JSWang:EPJB08_Quantum}
\begin{equation}
\Xi(\omega)=\text{Tr}[\boldsymbol{\Gamma}_{\text{R}}\boldsymbol{G}_{\text{RL}}^{\text{ret}}\boldsymbol{\Gamma}_{\text{L}}(\boldsymbol{G}_{\text{RL}}^{\text{ret}})^{\dagger}]\label{eq:CaroliFormula}
\end{equation}
 where $\boldsymbol{\Gamma}_{\text{L}}=i\boldsymbol{H}_{\text{L}}^{10}(\boldsymbol{g}_{\text{L},-}^{\text{ret}}-\boldsymbol{g}_{\text{L},-}^{\text{adv}})\boldsymbol{H}_{\text{L}}^{01}$
and $\boldsymbol{\Gamma}_{\text{R}}=i\boldsymbol{H}_{\text{R}}^{01}(\boldsymbol{g}_{\text{R},+}^{\text{ret}}-\boldsymbol{g}_{\text{R},+}^{\text{adv}})\boldsymbol{H}_{\text{R}}^{10}$.

\subsection{Phonon transmission, reflection and $S$-matrix}

From the Green's function $\boldsymbol{G}^{\text{ret}}$ in Eq.~(\ref{eq:FiniteGreensFunction}),
we can use the traditional AGF method to compute the phonon transmittance
$\Xi(\omega)$ which is the sum of the individual phonon transmission
coefficients.~\citep{ZHuang:JHT11_Modeling,ZYOng:PRB15_Efficient}
A more explicit connection to conventional scattering theory may be
made by noting that the transmission coefficients can be derived directly
from the diagonal elements of the transmission matrix,~\citep{DSFisher:PRB81_Relation}
which relates the amplitude of the incoming phonon flux to that of
the outgoing forward-scattered (or transmitted) phonon flux and is
computed numerically from $\boldsymbol{G}^{\text{ret}}$.~\citep{ZYOng:PRB15_Efficient}
However, this picture of the scattering process is incomplete as it
does not treat the amplitude of the \emph{backward}-scattered (or
reflected) phonons and the trajectories of the phonons reflected from
the interface. This suggests that a matrix analogous to the transmission
matrix is needed for the backward component of the scattered phonons.
To accomplish this, we introduce the reflection matrix and show how
it can be computed efficiently by building on the technical ideas
given in Ref.~\citep{ZYOng:PRB15_Efficient}. The reflection matrix
for each lead can then be combined with the transmission matrices
to form the \emph{$S$ matrix} that governs overall phonon transmission
and reflection at the interface. 

\subsubsection{Definition of transmission, absorption and reflection coefficients }

Before we proceed, we clarify some of the terminology used in the
following discussions. An \emph{incident} or ``incoming'' phonon is
one that has its group velocity pointing towards the interface and
corresponds to the asymptotically free ($t\rightarrow-\infty$) bulk
phonon state prior to scattering while an ``outgoing'' phonon is one
that has its group velocity pointing away from the interface and corresponds
to the asymptotically free ($t\rightarrow\infty$) bulk phonon state
after scattering. There are two types of outgoing phonons: (1) the
\emph{transmitted} or forward-scattered phonons on the other side
of the interface with a group velocity in the same direction as that
of the incident phonon and (2) the \emph{reflected} or backward-scattered
phonons on the same side of the interface but with a group velocity
opposite to that of the incident phonon. For example, an incoming
phonon in the left lead propagating towards the interface has a positive
group velocity. After colliding with the interface, the incoming phonon
is scattered into a range of outgoing phonon states, transmitted and
reflected, with a ``scattering amplitude'' and ``transition probability''
associated with each transition between the incoming phonon state
and an outgoing phonon state. 

We also use the transmission, absorption and reflection coefficients,
which can be obtained from sums of the relevant transition probabilities,
to characterize the loss and gain of energy by phonon channels. The
transmission coefficient $\Xi$ associated with each incoming phonon
channel is defined as the fraction of the energy flux \emph{lost}
by the incoming phonon channel across the interface to all the outgoing
phonon channels on the other side. The absorption coefficient $\xi$
associated with each outgoing phonon channel is defined as the fraction
of the energy flux gained by the outgoing phonon channel from all
the incoming phonon channels across the interface. Similarly, we can
also associate a reflection coefficient $\xi^{\prime}$ with each
outgoing phonon channel, which we define as the fraction of the energy
flux gained by the outgoing phonon channel from all the incoming channels
on the \emph{same} side of the interface. 

\subsubsection{Bloch matrices and bulk phonon eigenmodes}

The advanced and retarded Bloch matrices~\citep{TAndo:PRB91_Quantum,PAKhomyakov:PRB05_Conductance,ZYOng:PRB15_Efficient}
of the left and right lead, $\boldsymbol{F}_{\alpha}^{\text{adv/ret}}(+)$
and $\boldsymbol{F}_{\alpha}^{\text{adv/ret}}(-)$, describe their
bulk translational symmetry along the direction of the heat flux and
can be computed directly from the formulas: \begin{subequations}
\begin{equation}
\boldsymbol{F}_{\alpha}^{\text{adv/ret}}(+)=\boldsymbol{g}_{\alpha,+}^{\text{adv/ret}}\boldsymbol{H}_{\alpha}^{10}\label{eq:RightGoingBlochMatrix}
\end{equation}
\begin{equation}
\boldsymbol{F}_{\alpha}^{\text{adv/ret}}(-)^{-1}=\boldsymbol{g}_{\alpha,-}^{\text{adv/ret}}\boldsymbol{H}_{\alpha}^{01}\label{eq:LeftGoingBlochMatrix}
\end{equation}
\label{eq:BlochMatrices}\end{subequations} As pointed out in Ref.~\citep{ZYOng:PRB15_Efficient},
the bulk eigenmodes for the lead can be determined directly from the
Bloch matrices: \begin{subequations} 
\begin{equation}
\boldsymbol{F}_{\alpha}^{\text{adv/ret}}(+)\boldsymbol{U}_{\alpha}^{\text{adv/ret}}(+)=\boldsymbol{U}_{\alpha}^{\text{adv/ret}}(+)\boldsymbol{\Lambda}_{\alpha}^{\text{adv/ret}}(+)\label{eq:RightGoingModes}
\end{equation}
\begin{equation}
\boldsymbol{F}_{\alpha}^{\text{adv/ret}}(-)^{-1}\boldsymbol{U}_{\alpha}^{\text{adv/ret}}(-)=\boldsymbol{U}_{\alpha}^{\text{adv/ret}}(-)\boldsymbol{\Lambda}_{\alpha}^{\text{adv/ret}}(-)^{-1}\label{eq:LeftGoingModes}
\end{equation}
\label{eq:BlochMatrixEigenmodes}\end{subequations} where $\boldsymbol{U}_{\alpha}^{\text{ret}}(+)$
{[}$\boldsymbol{U}_{\alpha}^{\text{ret}}(-)${]} is a matrix with
its column vectors corresponding to the rightward-going (leftward-going)
extended or rightward (leftward) decaying evanescent modes and has
the form $\boldsymbol{U}_{\alpha}^{\text{ret}}=(\boldsymbol{e}_{1}\boldsymbol{e}_{2}\ldots\boldsymbol{e}_{N})$
where $\boldsymbol{e}_{n}$ is a normalized eigenvector of the Bloch
matrix in the $n$-th column of $\boldsymbol{U}_{\alpha}^{\text{ret}}$
{[}$\boldsymbol{U}_{\alpha}^{\text{ret}}(-)${]}. Similarly, $\boldsymbol{U}_{\alpha}^{\text{adv}}(-)$
{[}$\boldsymbol{U}_{\alpha}^{\text{adv}}(+)${]} is a matrix with
its column vectors corresponding to rightward-going (leftward-going)
extended or leftward (rightward) decaying evanescent modes. The matrix
$\boldsymbol{\Lambda}_{\alpha}^{\text{adv/ret}}(+)$ {[}$\boldsymbol{\Lambda}_{\alpha}^{\text{adv/ret}}(-)${]}
is a diagonal matrix with matrix elements of the form $e^{ik_{n}a}$
where $k_{n}$ is the phonon wave vector corresponding to the $n$-th
column eigenvector in $\boldsymbol{U}_{\alpha}^{\text{adv/ret}}(+)$
{[}$\boldsymbol{U}_{\alpha}^{\text{adv/ret}}(-)${]}. 

We note that because the Bloch matrices are not Hermitian, their eigenvectors
are not necessarily orthogonal. This can pose a problem~\citep{SSadasivam:PRB17_Phonon}
for transmission coefficient calculations when the eigenvectors have
the same $k$ and are degenerate. This issue can be simply resolved
by orthonormalizing the degenerate column eigenvectors in $\boldsymbol{U}_{\alpha}^{\text{adv/ret}}$
with a Gram-Schmidt procedure.~\citep{GArfken:Book95_Mathematical,CMWerneth:EJP10_Numerical}
The final piece of ingredient needed for the following phonon scattering
calculations is the diagonal velocity matrix~\citep{PAKhomyakov:PRB05_Conductance,JSWang:EPJB08_Quantum}
\begin{align}
\boldsymbol{V}_{\alpha}^{\text{adv/ret}}(+)= & \frac{ia_{\alpha}}{2\omega}[\boldsymbol{U}_{\alpha}^{\text{adv/ret}}(+)]^{\dagger}\boldsymbol{H}_{\alpha}^{01}[\boldsymbol{g}_{\alpha,+}^{\text{adv/ret}}-\nonumber \\
 & (\boldsymbol{g}_{\alpha,+}^{\text{ret/adv}})^{\dagger}]\boldsymbol{H}_{\alpha}^{10}\boldsymbol{U}_{\alpha}^{\text{adv/ret}}(+)\ ,\label{eq:RightGoingVelocityMatrix}
\end{align}
which has group velocities of the eigenvectors in $\boldsymbol{U}_{\alpha}^{\text{adv/ret}}(+)$
as its diagonal elements. Likewise, $\boldsymbol{V}_{\alpha}^{\text{adv/ret}}(-)$
is defined as
\begin{align}
\boldsymbol{V}_{\alpha}^{\text{adv/ret}}(-)= & -\frac{ia_{\alpha}}{2\omega}[\boldsymbol{U}_{\alpha}^{\text{adv/ret}}(-)]^{\dagger}\boldsymbol{H}_{\alpha}^{10}[\boldsymbol{g}_{\alpha,-}^{\text{adv/ret}}-\nonumber \\
 & (\boldsymbol{g}_{\alpha,-}^{\text{ret/adv}})^{\dagger}]\boldsymbol{H}_{\alpha}^{01}\boldsymbol{U}_{\alpha}^{\text{adv/ret}}(-)\ .\label{eq:LeftGoingVelocityMatrix}
\end{align}
For evanescent modes, the group velocity is always zero while for
propagating modes that contribute to the heat flux, the group velocity
is positive (negative) in $\boldsymbol{V}_{\alpha}^{\text{ret}}(+)$
and $\boldsymbol{V}_{\alpha}^{\text{adv}}(-)$ {[}$\boldsymbol{V}_{\alpha}^{\text{ret}}(-)$
and $\boldsymbol{V}_{\alpha}^{\text{adv}}(+)${]}. In addition, we
define the diagonal matrices $\widetilde{\boldsymbol{V}}_{\alpha}^{\text{adv/ret}}(+)$
and $\widetilde{\boldsymbol{V}}_{\alpha}^{\text{adv/ret}}(-)$ in
which their nonzero diagonal matrix elements are the inverse of those
of $\boldsymbol{V}_{\alpha}^{\text{adv/ret}}(+)$ and $\boldsymbol{V}_{\alpha}^{\text{adv/ret}}(-),$
respectively. For each lead, we can also define the diagonal matrices
\begin{subequations}
\begin{equation}
\boldsymbol{I}_{\alpha}^{\text{adv/ret}}(+)=\boldsymbol{V}_{\alpha}^{\text{adv/ret}}(+)\widetilde{\boldsymbol{V}}_{\alpha}^{\text{adv/ret}}(+)\label{eq:RightGoingIdentityMatrix}
\end{equation}
\begin{equation}
\boldsymbol{I}_{\alpha}^{\text{adv/ret}}(-)=\boldsymbol{V}_{\alpha}^{\text{adv/ret}}(-)\widetilde{\boldsymbol{V}}_{\alpha}^{\text{adv/ret}}(-)\label{eq:LeftGoingIdentityMatrix}
\end{equation}
\label{eq:AllGoingIdentityMatrices}\end{subequations} in which the
$n$-th diagonal element equals $1$ if the $n$-th column of $\boldsymbol{U}_{\alpha}^{\text{adv/ret}}(+)$
and $\boldsymbol{U}_{\alpha}^{\text{adv/ret}}(-)$ corresponds to
an extended mode and $0$ otherwise. Therefore, it follows from Eq.~(\ref{eq:AllGoingIdentityMatrices}
) that the number of rightward-going phonon channels $N_{\alpha}(+)$
and the number of leftward-going phonon channels $N_{\alpha}(-)$
are given by \begin{subequations}
\begin{equation}
N_{\alpha}(+)=\text{Tr}[\boldsymbol{I}_{\alpha}^{\text{ret}}(+)]=\text{Tr}[\boldsymbol{I}_{\alpha}^{\text{adv}}(-)]\label{eq:NumberRightGoingChannels}
\end{equation}
\begin{equation}
N_{\alpha}(-)=\text{Tr}[\boldsymbol{I}_{\alpha}^{\text{ret}}(-)]=\text{Tr}[\boldsymbol{I}_{\alpha}^{\text{adv}}(+)]\ .\label{eq:NumberLeftGoingChannels}
\end{equation}
\label{eq:AllNumberOfChannels}\end{subequations} 

\subsubsection{Phonon scattering: transmission}

Now, let us consider the scattering problem for an incoming phonon
from the left lead that is incident on the scattering region. In the
$n=0$ slice at the edge of the left lead, the motion can be decomposed
into two parts, i.e. 
\begin{equation}
\boldsymbol{c}_{0}=\boldsymbol{c}_{0}(+)+\boldsymbol{c}_{0}(-)\label{eq:Slice0Motion}
\end{equation}
where $\boldsymbol{c}_{0}(+)$ and $\boldsymbol{c}_{0}(-)$ respectively
represent the rightward-going (incident) and leftward-going (reflected)
components, while in the $n=2$ slice at the edge of the right lead,
the motion is given by 
\begin{equation}
\boldsymbol{c}_{2}=\boldsymbol{c}_{2}(+)\ ,\label{eq:TransmittedWave}
\end{equation}
where the  right-hand side represents a rightward-going (transmitted)
wave which can be a linear combination of bulk right-lead phonon modes
propagating away from the interface. Suppose the rightward-going component
in Eq.~(\ref{eq:Slice0Motion}) is a left-lead bulk phonon mode,
i.e. $\boldsymbol{c}_{0}(+)=\boldsymbol{u}_{\text{L},n}(k,\omega)$
where $n$ and $k$ are the phonon polarization index and wave vector,
respectively. Then, it can be shown~\citep{PAKhomyakov:PRB05_Conductance}
that the transmitted wave $\boldsymbol{c}_{2}(+)$ in the right lead
is related to the incident wave $\boldsymbol{c}_{0}(+)$ from the
right lead, via the expression 
\begin{equation}
\boldsymbol{c}_{2}=\boldsymbol{G}_{\text{RL}}^{\text{ret}}\boldsymbol{Q}_{\text{L}}\boldsymbol{u}_{\text{L},n}(k,\omega)\label{eq:TransmittedIncidentWave}
\end{equation}
where 
\begin{align}
\boldsymbol{Q}_{\alpha}= & (\omega^{2}+i\eta)\boldsymbol{I}_{\alpha}-\boldsymbol{H}_{\alpha}^{00}\nonumber \\
 & -\boldsymbol{H}_{\alpha}^{10}\boldsymbol{g}_{\alpha,-}^{\text{ret}}(\omega)\boldsymbol{H}_{\alpha}^{01}-\boldsymbol{H}_{\alpha}^{01}\boldsymbol{g}_{\alpha,+}^{\text{ret}}(\omega)\boldsymbol{H}_{\alpha}^{10}\label{eq:InverseBulkGreensFunction}
\end{align}
and $\boldsymbol{Q}_{\alpha}^{-1}$ is the bulk Green's function of
the $\alpha$ lead. The expression in Eq.~(\ref{eq:TransmittedIncidentWave})
can be expressed as a linear combination of transmitted right-lead
phonon modes $\boldsymbol{u}_{\text{R},m}(k_{m},\omega)$, i.e. $\boldsymbol{c}_{2}=\sum_{m}\boldsymbol{u}_{\text{R},m}(k_{m},\omega)\tau_{mn}$,
where $\tau_{mn}$ is the linear coefficient and forms the matrix
elements of the transmission matrix $\boldsymbol{\tau}$, where
\begin{equation}
\boldsymbol{\tau}=[\boldsymbol{U}_{\text{R}}^{\text{ret}}(+)]^{-1}\boldsymbol{G}_{\text{RL}}^{\text{ret}}\boldsymbol{Q}_{\text{L}}\boldsymbol{U}_{\text{L}}^{\text{ret}}(+)\ .\label{eq:taumatrix}
\end{equation}
The flux-normalized transmission matrix is $\boldsymbol{t}_{\text{RL}}=[\boldsymbol{V}_{\text{R}}^{\text{ret}}(+)]^{\nicefrac{1}{2}}\boldsymbol{\tau}[\widetilde{\boldsymbol{V}}_{\text{L}}^{\text{ret}}(+)]^{\nicefrac{1}{2}}$,
which we can rewrite as~\citep{ZYOng:PRB15_Efficient} 
\begin{align}
\boldsymbol{t}_{\text{RL}}= & \frac{2i\omega}{\sqrt{a_{\text{R}}a_{\text{L}}}}[\boldsymbol{V}_{\text{R}}^{\text{ret}}(+)]^{\nicefrac{1}{2}}[\boldsymbol{U}_{\text{R}}^{\text{ret}}(+)]^{-1}\nonumber \\
 & \times\boldsymbol{G}_{\text{RL}}^{\text{ret}}[\boldsymbol{U}_{\text{L}}^{\text{adv}}(-)^{\dagger}]^{-1}[\boldsymbol{V}_{\text{L}}^{\text{adv}}(-)]^{\nicefrac{1}{2}}\ .\label{eq:tmatrix_RL}
\end{align}
Each row of $\boldsymbol{t}_{\text{RL}}$ corresponds to either a
transmitted right-lead extended or evanescent mode. For an outgoing
evanescent mode, the row elements and group velocity, given by the
diagonal element of $\boldsymbol{V}_{\text{R}}^{\text{ret}}(+)$,
are zero. Conversely, each column of of $\boldsymbol{t}_{\text{RL}}$
corresponds to either an incident left-lead extended or evanescent
mode, and the column elements and group velocity of the evanescent
modes, given by the diagonal element of $\boldsymbol{V}_{\text{L}}^{\text{adv}}(-)$,
are zero. If the $m$-th row and $n$-th column of $\boldsymbol{t}_{\text{RL}}$correspond
to extended transmitted and incident modes, then $|[\boldsymbol{t}_{\text{RL}}]_{mn}|^{2}$
gives us the probability that incident left-lead phonon is transmitted
across the interface into the right-lead phonon. Similarly, we can
define the flux-normalized transmission matrix for phonon transmission
from the right to the left lead:

\begin{align}
\boldsymbol{t}_{\text{LR}}= & \frac{2i\omega}{\sqrt{a_{\text{L}}a_{\text{R}}}}[\boldsymbol{V}_{\text{L}}^{\text{ret}}(-)]^{\nicefrac{1}{2}}[\boldsymbol{U}_{\text{L}}^{\text{ret}}(-)]^{-1}\nonumber \\
 & \times\boldsymbol{G}_{\text{LR}}^{\text{ret}}[\boldsymbol{U}_{\text{R}}^{\text{adv}}(+)^{\dagger}]^{-1}[\boldsymbol{V}_{\text{R}}^{\text{adv}}(+)]^{\nicefrac{1}{2}}\ .\label{eq:tmatrix_LR}
\end{align}

\subsubsection{Phonon scattering: reflection}

Like in Eq.~(\ref{eq:TransmittedIncidentWave}), we can describe
the motion in slice 0 in terms of the incident wave, i.e. $\boldsymbol{c}_{0}=\boldsymbol{G}_{\text{L}}^{\text{ret}}\boldsymbol{Q}_{\text{L}}\boldsymbol{u}_{\text{L},n}(k,\omega)$.
It follows that the reflected component is $\boldsymbol{c}_{0}(-)=\boldsymbol{c}_{0}-\boldsymbol{c}_{0}(+)=(\boldsymbol{G}_{\text{L}}^{\text{ret}}-\boldsymbol{Q}_{\text{L}}^{-1})\boldsymbol{Q}_{\text{L}}\boldsymbol{u}_{\text{L},n}(k,\omega)$.
Therefore, the flux-normalized reflection matrix, which gives the
scattering amplitude between leftward-going (reflected) and rightward-going
(incident) states in the left lead, can be defined as:
\begin{align}
\boldsymbol{r}_{\text{LL}}= & \frac{2i\omega}{a_{\text{L}}}[\boldsymbol{V}_{\text{L}}^{\text{ret}}(-)]^{\nicefrac{1}{2}}[\boldsymbol{U}_{\text{L}}^{\text{ret}}(-)]^{-1}\nonumber \\
 & \times(\boldsymbol{G}_{\text{L}}^{\text{ret}}-\boldsymbol{Q}_{\text{L}}^{-1})[\boldsymbol{U}_{\text{L}}^{\text{adv}}(-)^{\dagger}]^{-1}[\boldsymbol{V}_{\text{L}}^{\text{adv}}(-)]^{\nicefrac{1}{2}}\ .\label{eq:rmatrix_LL}
\end{align}
The corresponding expression for phonon reflection in the right lead
can be similarly defined as 
\begin{align}
\boldsymbol{r}_{\text{RR}}= & \frac{2i\omega}{a_{\text{R}}}[\boldsymbol{V}_{\text{R}}^{\text{ret}}(+)]^{\nicefrac{1}{2}}[\boldsymbol{U}_{\text{R}}^{\text{ret}}(+)]^{-1}\nonumber \\
 & \times(\boldsymbol{G}_{\text{R}}^{\text{ret}}-\boldsymbol{Q}_{\text{R}}^{-1})[\boldsymbol{U}_{\text{R}}^{\text{adv}}(+)^{\dagger}]^{-1}[\boldsymbol{V}_{\text{R}}^{\text{adv}}(+)]^{\nicefrac{1}{2}}\ ,\label{eq:rmatrix_RR}
\end{align}
which gives the scattering amplitude between rightward-going (reflected)
and leftward-going (incident) states in the right lead.

\subsubsection{Phonon transmission and reflection matrices }

Given Eqs.~(\ref{eq:tmatrix_RL}) to (\ref{eq:rmatrix_RR}), we can
construct the rationalized smaller matrices $\bar{\boldsymbol{t}}_{\text{RL}}$,
$\bar{\boldsymbol{t}}_{\text{LR}}$, $\bar{\boldsymbol{r}}_{\text{LL}}$
and $\bar{\boldsymbol{r}}_{\text{RR}}$ from $\boldsymbol{t}_{\text{RL}}$,
$\boldsymbol{t}_{\text{LR}}$, $\boldsymbol{r}_{\text{LL}}$ and $\boldsymbol{r}_{\text{RR}}$
by deleting the matrix rows and columns corresponding to evanescent
states. This is done numerically by inspecting each diagonal element
of $\boldsymbol{I}_{\alpha}^{\text{adv/ret}}(\pm)$ of Eq.~(\ref{eq:AllGoingIdentityMatrices}),
which is either equal to 0 (evanescent) or 1 (extended), and removing
the corresponding columns or rows when $[\boldsymbol{I}_{\alpha}^{\text{adv/ret}}(\pm)]_{nn}=0$.
For example, to find $\bar{\boldsymbol{t}}_{\text{RL}}$, we inspect
$\boldsymbol{I}_{\text{R}}^{\text{ret}}(+)$ for row deletion and
$\boldsymbol{I}_{\text{L}}^{\text{adv}}(-)$ for column deletion in
$\boldsymbol{t}_{\text{RL}}$. Hence, $\bar{\boldsymbol{t}}_{\text{RL}}$
is an $N_{\text{R}}(+)\times N_{\text{L}}(+)$ matrix. Similarly,
we can also define the rationalized smaller matrices $\bar{\boldsymbol{\Lambda}}_{\alpha}^{\text{adv/ret}}(+)$
by deleting the rows and columns associated with evanescent modes
from $\boldsymbol{\Lambda}_{\alpha}^{\text{adv/ret}}(\pm)$ in Eq.~(\ref{eq:BlochMatrixEigenmodes}). 

The\emph{ transmission coefficient} of the $n$-th \emph{incoming}
phonon channel in the left lead is defined as the $n$-th diagonal
element of $\bar{\boldsymbol{t}}_{\text{RL}}^{\dagger}\bar{\boldsymbol{t}}_{\text{RL}}$,
i.e.
\begin{equation}
\Xi_{\text{L},n}=[\bar{\boldsymbol{t}}_{\text{RL}}^{\dagger}\bar{\boldsymbol{t}}_{\text{RL}}]_{nn}\ ,\label{eq:IncomingTransmitCoeff}
\end{equation}
which is equal to the fraction of its energy flux transmitted across
the interface, and its wave vector $k_{n}$ can be determined from
$[\bar{\boldsymbol{\Lambda}}_{\text{L}}^{\text{adv}}(-)]_{nn}=e^{ik_{n}a_{\text{L}}}$
or $k_{n}=\frac{1}{a_{\text{L}}}\cos^{-1}\text{Re}[\bar{\boldsymbol{\Lambda}}_{\text{L}}^{\text{adv}}(-)]_{nn}$. For
the reflected modes, the \emph{reflection coefficient} of the $m$-th
outgoing leftward-going mode in the left lead is given by the $m$-th
diagonal element of $\bar{\boldsymbol{r}}_{\text{LL}}\bar{\boldsymbol{r}}_{\text{LL}}^{\dagger}$,
i.e.
\begin{equation}
\xi_{\text{L},m}^{\prime}=[\bar{\boldsymbol{r}}_{\text{LL}}\bar{\boldsymbol{r}}_{\text{LL}}^{\dagger}]_{mm}\ ,\label{eq:OutgoingReflectCoeff}
\end{equation}
with its phonon wave vector $k_{m}$ given by $k_{m}=\frac{1}{a_{\text{L}}}\cos^{-1}\text{Re}[\bar{\boldsymbol{\Lambda}}_{\text{L}}^{\text{ret}}(-)]_{mm}$,
while the \emph{absorption coefficient} of the $l$-th outgoing rightward-going
mode in the right lead is given by the $l$-th diagonal element of
$\bar{\boldsymbol{t}}_{\text{RL}}\bar{\boldsymbol{t}}_{\text{RL}}^{\dagger}$,
i.e.
\begin{equation}
\xi_{\text{R},l}=[\bar{\boldsymbol{t}}_{\text{RL}}\bar{\boldsymbol{t}}_{\text{RL}}^{\dagger}]_{ll}\ ,\label{eq:OutgoingTransmitCoeff}
\end{equation}
with its phonon wave vector $k_{l}$ given by $k_{l}=\frac{1}{a_{\text{R}}}\cos^{-1}\text{Re}[\bar{\boldsymbol{\Lambda}}_{\text{R}}^{\text{ret}}(+)]_{ll}$. 

The transmission coefficient for the $n$-th incoming phonon channel
in the right lead ($\Xi_{\text{R},n}$), the absorption coefficient
of the $l$-th outgoing phonon channel in the left lead ($\xi_{\text{L},l}$
) and the reflection coefficient of the $m$-th outgoing phonon channel
in the right lead ($\xi_{\text{R},m}^{\prime}$) can be similarly
defined like in Eqs.~(\ref{eq:IncomingTransmitCoeff}) to (\ref{eq:OutgoingTransmitCoeff}),
and their formulas are summarized in Table~\ref{tab:PhononFormulae}.
It should also be noted that for $\alpha=\text{L},\text{R}$,
\begin{equation}
\xi_{\alpha,m}+\xi_{\alpha,m}^{\prime}=1\label{eq:SumEnergyFluxFraction}
\end{equation}
which physically means that the sum of the energy flux fractions from
absorption and reflection equals unity, consistent with the conservation
of energy. In addition, we remark that the phonon transmittance can
be expressed as the sum of the transmission {[}Eq.~(\ref{eq:IncomingTransmission}){]}
or absorption {[}Eq.~(\ref{eq:OutgoingTransmission}){]} coefficients
of either lead, i.e. \begin{subequations}
\begin{align}
\Xi(\omega) & =\sum_{n=1}^{N_{\text{L}}(+)}\Xi_{\text{L},n}=\sum_{m=1}^{N_{\text{R}}(-)}\Xi_{\text{R},m}\ \label{eq:IncomingTransmission}\\
 & =\sum_{n=1}^{N_{\text{L}}(-)}\xi_{\text{L},n}=\sum_{m=1}^{N_{\text{R}}(+)}\xi_{\text{R},m}\ .\label{eq:OutgoingTransmission}
\end{align}

\label{eq:AllPhononTransmittace}\end{subequations}

\subsubsection{Phonon scattering specularity}

With our method, the phonon scattering specularity parameter, which
measures the `smoothness' of a surface, can be extracted directly
from the reflection matrices $\bar{\boldsymbol{r}}_{\text{LL}}$ and
$\bar{\boldsymbol{r}}_{\text{RR}}$. Here, we discuss briefly the
meaning of phonon specularity and how it is computed in the AGF-based
$S$-matrix approach. In Ref.~\citep{JZiman:Book60_Electrons}, the
specularity parameter is simply defined as the proportion of the intensity
of the incident wave that remains in the outgoing wave in the specular
direction, with the effects of polarization conversion ignored and
the rest of the intensity assumed to be redistributed equally in all
directions. In our $S$-matrix approach, we adopt a similar definition
for atomistic phonon scattering specularity $\mathcal{P}$ by taking
it to be the intensity proportion that is scattered to the specularly
reflected outgoing channel, which we define as the outgoing phonon
channel with the longitudinal wave vector $k_{\bar{n}}=-k_{n}$ and
of the same polarization. However, we caution that this definition
of specularity does not necessarily imply that the remainder is equally
distributed in the rest of the outgoing channels, i.e. the absence
of specularity does not correspond to diffusive scattering. 

In the case of \emph{total} phonon reflection in the left lead, the
\emph{specularity parameter} $\mathcal{P}_{\text{L}}(k_{n})$ for
the incoming left-lead phonon at $k_{n}$ is determined by its transition
probability to the outgoing phonon channel at $k_{\bar{n}}$, i.e.
\begin{equation}
\mathcal{P}_{\text{L}}(k_{n})=|[\bar{\boldsymbol{r}}_{\text{LL}}^{\dagger}]_{n\bar{n}}|^{2}\ .\label{eq:LeftLeadSpecularityParameterDefn}
\end{equation}
The expression in Eq.~(\ref{eq:LeftLeadSpecularityParameterDefn})
satisfies the requirement that $\mathcal{P}=1$ for fully specular
reflection and in the limit that the number of channels goes to infinity,
$\mathcal{P}=0$ for fully diffusive scattering.~\citep{JZiman:Book60_Electrons}
In the more general case of \emph{partial} phonon reflection and transmission
at an interface, the specularity parameter for the mode at $k_{n}$
in Eq.~(\ref{eq:LeftLeadSpecularityParameterDefn}) has to be normalized
by the overall probability of its phonon reflection, giving us
\begin{equation}
\mathcal{P}_{\text{L}}(k_{n})=\frac{|[\bar{\boldsymbol{r}}_{\text{LL}}^{\dagger}]_{n\bar{n}}|^{2}}{\sum_{m}|[\bar{\boldsymbol{r}}_{\text{LL}}^{\dagger}]_{nm}|^{2}}=\frac{|[\bar{\boldsymbol{r}}_{\text{LL}}^{\dagger}]_{n\bar{n}}|^{2}}{[\bar{\boldsymbol{r}}_{\text{LL}}^{\dagger}\bar{\boldsymbol{r}}_{\text{LL}}]_{nn}}\ .\label{eq:NormalizedLeftLeadSpecularityParameterDefn}
\end{equation}
Similarly, the specularity parameter for an incoming right-lead phonon
with the wave vector $k_{m}$ is $\mathcal{P}_{\text{R}}(k_{m})=|[\bar{\boldsymbol{r}}_{\text{RR}}^{\dagger}]_{m\bar{m}}|^{2}/[\bar{\boldsymbol{r}}_{\text{RR}}^{\dagger}\bar{\boldsymbol{r}}_{\text{RR}}]_{mm}\ .$

\begin{table*}
\begin{tabular}{|l|l|l|}
\hline 
Variable & Formula & Phonon wave vector\tabularnewline
\hline 
\hline 
Incoming left-lead phonon transmission coefficient & $\Xi_{\text{L},n}=[\bar{\boldsymbol{t}}_{\text{RL}}^{\dagger}\bar{\boldsymbol{t}}_{\text{RL}}]_{nn}$ & $k_{n}=\frac{1}{a_{\text{L}}}\cos^{-1}\text{Re}[\bar{\boldsymbol{\Lambda}}_{\text{L}}^{\text{adv}}(-)]_{nn}$\tabularnewline
\hline 
Outgoing right-lead phonon absorption coefficient & $\xi_{\text{R},n}=[\bar{\boldsymbol{t}}_{\text{RL}}\bar{\boldsymbol{t}}_{\text{RL}}^{\dagger}]_{nn}$ & $k_{n}=\frac{1}{a_{\text{R}}}\cos^{-1}\text{Re}[\bar{\boldsymbol{\Lambda}}_{\text{R}}^{\text{ret}}(+)]_{nn}$\tabularnewline
\hline 
Outgoing left-lead phonon reflection coefficient & $\xi_{\text{L},n}^{\prime}=[\bar{\boldsymbol{r}}_{\text{LL}}\bar{\boldsymbol{r}}_{\text{LL}}^{\dagger}]_{nn}$ & $k_{n}=\frac{1}{a_{\text{L}}}\cos^{-1}\text{Re}[\bar{\boldsymbol{\Lambda}}_{\text{L}}^{\text{ret}}(-)]_{nn}$\tabularnewline
\hline 
Incoming right-lead phonon transmission coefficient & $\Xi_{\text{R},n}=[\bar{\boldsymbol{t}}_{\text{LR}}^{\dagger}\bar{\boldsymbol{t}}_{\text{LR}}]_{nn}$ & $k_{n}=\frac{1}{a_{\text{R}}}\cos^{-1}\text{Re}[\bar{\boldsymbol{\Lambda}}_{\text{R}}^{\text{adv}}(+)]_{nn}$\tabularnewline
\hline 
Outgoing left-lead phonon absorption coefficient & $\xi_{\text{L},n}=[\bar{\boldsymbol{t}}_{\text{LR}}\bar{\boldsymbol{t}}_{\text{LR}}^{\dagger}]_{nn}$ & $k_{n}=\frac{1}{a_{\text{L}}}\cos^{-1}\text{Re}[\bar{\boldsymbol{\Lambda}}_{\text{L}}^{\text{ret}}(-)]_{nn}$\tabularnewline
\hline 
Outgoing right-lead phonon reflection coefficient & $\xi_{\text{R},n}^{\prime}=[\bar{\boldsymbol{r}}_{\text{RR}}\bar{\boldsymbol{r}}_{\text{RR}}^{\dagger}]_{nn}$ & $k_{n}=\frac{1}{a_{\text{R}}}\cos^{-1}\text{Re}[\bar{\boldsymbol{\Lambda}}_{\text{R}}^{\text{ret}}(+)]_{nn}$\tabularnewline
\hline 
\end{tabular}

\caption{Summary of formulas for the phonon mode transmission, absorption and
reflection coefficients. The term `incoming' describes a phonon moving
\emph{towards} the interface while `outgoing' refers to phonons moving
\emph{away} from the interface. }
\label{tab:PhononFormulae}
\end{table*}

\subsubsection{$S$-matrix description of phonon scattering}

Given $\bar{\boldsymbol{t}}_{\text{RL}}$, $\bar{\boldsymbol{t}}_{\text{LR}}$,
$\bar{\boldsymbol{r}}_{\text{LL}}$ and $\bar{\boldsymbol{r}}_{\text{RR}}$,
we can define the $S$ matrix
\begin{equation}
\boldsymbol{S}=\left(\begin{array}{cc}
\bar{\boldsymbol{r}}_{\text{LL}} & \bar{\boldsymbol{t}}_{\text{LR}}\\
\bar{\boldsymbol{t}}_{\text{RL}} & \bar{\boldsymbol{r}}_{\text{RR}}
\end{array}\right)\ ,\label{eq:SMatrix}
\end{equation}
which connects the amplitudes of the scattered (reflected and transmitted)
bulk phonons to the incident bulk phonons and is unitary if the system
possesses time-reversal symmetry, i.e. $\boldsymbol{S}\boldsymbol{S}^{\dagger}=\boldsymbol{S}^{\dagger}\boldsymbol{S}=\boldsymbol{I}_{p}$
where $\boldsymbol{I}_{p}$ is an identity matrix of the same size
as $\boldsymbol{S}$. The unitarity of $\boldsymbol{S}$ allows us
to derive several identities involving $\bar{\boldsymbol{t}}_{\text{RL}}$,
$\bar{\boldsymbol{t}}_{\text{LR}}$, $\bar{\boldsymbol{r}}_{\text{LL}}$
and $\bar{\boldsymbol{r}}_{\text{RR}}$. Equations~(\ref{eq:AllNumberOfChannels})
and (\ref{eq:SMatrix}) imply that 
\begin{equation}
N_{\text{L}}(+)+N_{\text{R}}(-)=N_{\text{L}}(-)+N_{\text{R}}(+)\ ,\label{eq:IncomingOutgoingChannelEquality}
\end{equation}
i.e. the total number of incoming phonon channels is equal to the
total number of outgoing phonon channels. It follows from Eqs.~(\ref{eq:SMatrix})
and (\ref{eq:SumEnergyFluxFraction}) that
\begin{equation}
\left\{ \begin{array}{c}
N_{\text{L}}(+)\\
N_{\text{L}}(-)
\end{array}\right\} =\left\{ \begin{array}{c}
\text{Tr}(\bar{\boldsymbol{r}}_{\text{LL}}^{\dagger}\bar{\boldsymbol{r}}_{\text{LL}}+\bar{\boldsymbol{t}}_{\text{RL}}^{\dagger}\bar{\boldsymbol{t}}_{\text{RL}})\\
\text{Tr}(\bar{\boldsymbol{r}}_{\text{LL}}\bar{\boldsymbol{r}}_{\text{LL}}^{\dagger}+\bar{\boldsymbol{t}}_{\text{LR}}\bar{\boldsymbol{t}}_{\text{LR}}^{\dagger})
\end{array}\right\} \label{eq:LeftLeadChannelCount}
\end{equation}
and $N_{\text{L}}(+)=N_{\text{L}}(-)$, i.e., the number of leftward-going
bulk phonon channels is equal to the number of rightward-going bulk
phonon channels in the left lead. Similarly, we also have 
\begin{equation}
\left\{ \begin{array}{c}
N_{\text{R}}(-)\\
N_{\text{R}}(+)
\end{array}\right\} =\left\{ \begin{array}{c}
\text{Tr}(\bar{\boldsymbol{r}}_{\text{RR}}^{\dagger}\bar{\boldsymbol{r}}_{\text{RR}}+\bar{\boldsymbol{t}}_{\text{LR}}^{\dagger}\bar{\boldsymbol{t}}_{\text{LR}})\\
\text{Tr}(\bar{\boldsymbol{r}}_{\text{RR}}\bar{\boldsymbol{r}}_{\text{RR}}^{\dagger}+\bar{\boldsymbol{t}}_{\text{RL}}\bar{\boldsymbol{t}}_{\text{RL}}^{\dagger})
\end{array}\right\} \label{eq:RightLeadChannelCount}
\end{equation}
and $N_{\text{R}}(-)=N_{\text{R}}(+)$. Equations~(\ref{eq:LeftLeadChannelCount})
and (\ref{eq:RightLeadChannelCount}) also allow us to establish the
general \emph{reciprocity relationship},~\citep{AMaznev:WM13_Reciprocity}
\begin{equation}
\text{Tr}(\bar{\boldsymbol{t}}_{\text{RL}}\bar{\boldsymbol{t}}_{\text{RL}}^{\dagger})=\text{Tr}(\bar{\boldsymbol{t}}_{\text{LR}}\bar{\boldsymbol{t}}_{\text{LR}}^{\dagger})\ ,\label{eq:Reciprocity}
\end{equation}
or that the total rightward-going phonon transmission is equal to
total leftward-going phonon transmission. They also imply that the
phonon transmittance is bounded by the finite number of channels,
i.e., $\Xi(\omega)\leq\text{min}\left(N_{\text{L}}(+),N_{\text{R}}(-)\right)\ .$

\section{Example with carbon nanotube junction}

We illustrate the method by simulating phonon scattering at the armchair
junction between two isotopically different but structurally identical
(8,8) carbon nanotubes, as can be seen in Fig.~\ref{fig:CNTPhononPlot}(a),
with the left one (`CNT-12') consisting of $^{12}$C atoms and the
right one (`CNT-24') of $^{24}$C atoms which have twice the mass
of $^{12}$C atoms. The greater atomic mass of the $^{24}$C atom
doubles the mass density of CNT-24 and hence rescales its phonon frequencies
by a factor of $\frac{1}{\sqrt{2}}$, introducing a difference in
the polarization and distribution of phonon channels on either side
of the junction at each frequency $\omega$. However, the phonon dispersion
($\omega$ vs. $k$) curves in CNT-24 are identical in shape to those
of CNT-12 apart from the difference in frequency scaling. Thus, each
phonon branch or `subband' in CNT-12, which depends on polarization
and angular symmetry,~\citep{EDobardzic:PRB03_Single} has a unique
image subband in CNT-24 and as we shall show later, this simplifies
our analysis of the polarization dependence of phonon scattering.
Although $^{24}$C atoms do not exist, this fictitious system is sufficiently
realistic to contain the essential physics of phonon scattering by
an interface as well as to illustrate key concepts introduced in the
previous section. 

\subsection{Calculation details}

We build the carbon nanotube (CNT) and optimize its structure in GULP~\citep{JGale:MolSim03_gulp}
using the Tersoff potential~\citep{JTersoff:PRL88_Empirical} parameters
from Ref.~\citep{LLindsay:PRB10_Optimized}. The force-constant matrices
for the left and right leads ($\boldsymbol{H}_{\text{L}}^{00}$, $\boldsymbol{H}_{\text{L}}^{01}$,
$\boldsymbol{H}_{\text{R}}^{00}$ and $\boldsymbol{H}_{\text{R}}^{01}$)
are also computed in GULP. In our CNT structure, the interatomic interactions
are sufficiently short-range so that the \emph{primitive} unit cells
correspond to the individual slices in our AGF calculation. At each
frequency ($\omega$) point, we use the force-constant matrices to
find the surface Green's function $\boldsymbol{g}_{\text{R},+}^{\text{ret}}$
and $\boldsymbol{g}_{\text{L},-}^{\text{ret}}$, from which we determine
$\mathbf{H}^{\prime}$ and $\boldsymbol{G}^{\text{ret}}$ using Eqs.~(\ref{eq:ProjectedForceConstantMatrix})
and (\ref{eq:FiniteGreensFunction}). Using Eqs.~(\ref{eq:BlochMatrices})
and (\ref{eq:BlochMatrixEigenmodes}), we also calculate the incoming
phonon modes $\boldsymbol{U}_{\text{L}}^{\text{adv}}(-)$ and $\boldsymbol{U}_{\text{R}}^{\text{adv}}(+)$
and the outgoing phonon modes $\boldsymbol{U}_{\text{L}}^{\text{ret}}(-)$
and $\boldsymbol{U}_{\text{R}}^{\text{ret}}(+)$ as well as their
associated velocity matrices,$\boldsymbol{V}_{\text{L}}^{\text{adv}}(-)$,
$\boldsymbol{V}_{\text{R}}^{\text{adv}}(+)$, $\boldsymbol{V}_{\text{L}}^{\text{ret}}(-)$
and $\boldsymbol{V}_{\text{R}}^{\text{ret}}(+)$. The surface Green's
functions $\boldsymbol{g}_{\text{R},-}^{\text{ret}}$ and $\boldsymbol{g}_{\text{L},+}^{\text{ret}}$
are also computed and combined with $\boldsymbol{g}_{\text{R},+}^{\text{ret}}$
and $\boldsymbol{g}_{\text{L},-}^{\text{ret}}$ to find $\boldsymbol{Q}_{\text{R}}$
and $\boldsymbol{Q}_{\text{L}}$. Finally, these matrix variables
are collected and used to compute the transmission and reflection
matrices ($\boldsymbol{t}_{\text{RL}}$, $\boldsymbol{t}_{\text{LR}}$,
$\boldsymbol{r}_{\text{LL}}$ and $\boldsymbol{r}_{\text{RR}}$ )
in Eqs.~(\ref{eq:tmatrix_RL}) to (\ref{eq:rmatrix_RR}). We then
eliminate the non-physical matrix rows and columns from them to obtain
$\bar{\boldsymbol{t}}_{\text{RL}}$, $\bar{\boldsymbol{t}}_{\text{LR}}$,
$\bar{\boldsymbol{r}}_{\text{LL}}$ and $\bar{\boldsymbol{r}}_{\text{RR}}$
which constitute the $S$ matrix in Eq.~(\ref{eq:SMatrix}). The
transmission, absorption, and reflection coefficients of the phonon
channels for each CNT are computed, using Eqs.~(\ref{eq:IncomingTransmitCoeff})
to (\ref{eq:OutgoingTransmitCoeff}).

\subsection{Transmission, absorption and reflection coefficients}

\begin{figure}

\includegraphics[scale=0.67]{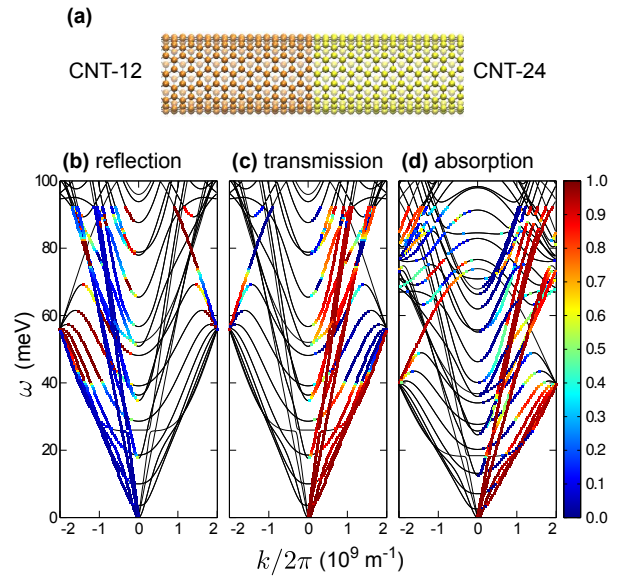}

\caption{\textbf{(a)} Plot of the armchair junction between two isotopically
different carbon nanotubes (CNT's). The left CNT (`CNT-12') is constructed
from $^{12}$C atoms while the right CNT (`CNT-24') has $^{24}$C
atoms. Phonon momentum and polarization-resolved plot of \textbf{(b)}
left-lead reflection coefficients, \textbf{(c)} left-lead transmission
coefficients, and \textbf{(d)} right-lead absorption coefficients
for phonon transmission from CNT-12 to CNT-24. }

\label{fig:CNTPhononPlot}
\end{figure}

We analyze the distribution of the transmission, absorption and reflection
coefficients for the incident phonon flux from CNT-12 to CNT-24. Figure~\ref{fig:CNTPhononPlot}(b)
shows the reflection coefficient distribution {[}$\xi_{\text{L},n}^{\prime}$
for $n=1,\ldots,N_{\text{L}}(-)${]} for the outgoing leftward-going
phonon modes while Fig.~\ref{fig:CNTPhononPlot}(c) shows the transmission
coefficient distribution {[}$\Xi_{\text{L},n}$ for $n=1,\ldots,N_{\text{L}}(+)${]}
for the incoming rightward-going phonon modes in CNT-12. On the other
side of the interface, the absorption coefficient distribution {[}$\xi_{\text{R},n}$
for $n=1,\ldots,N_{\text{R}}(+)${]} for the outgoing rightward-going
phonon modes in CNT-24 is shown in Fig.~\ref{fig:CNTPhononPlot}(d).
We also plot the phonon dispersion curves for CNT-12 and CNT-24 in
Fig.~\ref{fig:CNTPhononPlot} over the frequency range between 0
and 100 meV, with the individual phonon branches~\citep{MDresselhaus:AdvPhys00_Phonons}
clearly visible. In each spectrum, we note that only half of the points
on the dispersion curves contribute to the transmission or absorption/reflection
because half of the modes are either leftward or rightward-going.
Thus, only half of the phonon channels can contribute to the phonon
transmission or reflection at any frequency. 

Figure~\ref{fig:CNTPhononPlot}(c) shows that at low frequencies
($\omega<20$ meV), the transmission coefficients ($\Xi_{\text{L},n}$)
of all the incoming phonon modes are very close to unity, i.e. the
phonon modes in CNT-12 are nearly perfectly transmitted across the
interface. Conversely, the reflection coefficients ($\xi_{\text{L},n}^{\prime}$)
of the corresponding outgoing phonon modes in Fig.~\ref{fig:CNTPhononPlot}(b)
are close to zero at low frequencies. A comparison of Figs.~\ref{fig:CNTPhononPlot}(b)
and (c) shows that each reflected phonon mode at $k_{i}$ with a reflection
coefficient of $\xi_{\text{L},i}^{\prime}$ in Fig.~\ref{fig:CNTPhononPlot}(b)
corresponds symmetrically to a transmitted phonon mode at $k_{j}=-k_{i}$
with a transmission coefficient of $\Xi_{\text{L},j}=1-\xi_{\text{L},i}^{\prime}$
in Fig.~\ref{fig:CNTPhononPlot}(c). In CNT-24 {[}Fig.~\ref{fig:CNTPhononPlot}(d){]},
the absorption coefficient spectrum ($\xi_{\text{R},n}$) for the
outgoing phonon modes reveals that many of the rightward-going phonon
channels have an absorption coefficient close to zero even at low
frequencies although others have an absorption coefficient close to
unity, indicating that there are preferred outgoing channels and subbands
for phonon absorption. The presence of these $\xi_{\text{R},n}\sim0$
channels is because at the same frequency ($\omega$), there are generally
more phonon channels in CNT-24 than in CNT-12 and the phonon flux
at the interface is thus limited by the transmission bottleneck through
the fewer incoming phonon channels in CNT-12. The absorption coefficients
also tend to be lower for outgoing phonon modes nearer the phonon
subband edges and with a lower group velocity ($v=\frac{\partial\omega}{\partial k}$). 

\subsection{Transition probabilities of scattering processes}

In our analysis of the absorption spectrum in Fig.~\ref{fig:CNTPhononPlot}(d),
we find that energy is preferentially transmitted to some phonon subbands,
suggesting that transitions between phonon channels associated with
certain subbands are dominant. To elucidate the role of the subbands
in phonon scattering, we use our method to determine and analyze the
transition probabilities between different bulk phonon channels. We
analyze two sets of scattering processes, with the first corresponding
to an incoming phonon channel at $k_{1}$ in the left lead (CNT-12)
and the second to an incoming phonon channel at $\bar{k}_{3}$ in
the right lead (CNT-24), at $\omega=39.5$ meV. Here and in our subsequent
discussion of the scattering simulation results, to represent a phonon
wave vector of equal magnitude but directionally opposite to $k_{i}$,
we write a bar over the latter, i.e. $\bar{k}_{i}=-k_{i}$; the corresponding
integer index for $\bar{k}_{i}$ is written as $\bar{i}$. The transition
probabilities $P(k\rightarrow k^{\prime})$ for all available incoming
and outgoing phonon channels are computed from the square of the scattering
amplitudes determined from the matrix elements of $\bar{\boldsymbol{t}}_{\text{RL}}$,
$\bar{\boldsymbol{t}}_{\text{LR}}$, $\bar{\boldsymbol{r}}_{\text{LL}}$
and $\bar{\boldsymbol{r}}_{\text{RR}}$.

\subsubsection{Incoming phonon channel at $k_{1}$ in CNT-12~\label{subsec:Incoming_phonon_channel}}

Figures~\ref{fig:PhononTransitions}(a) and \ref{fig:PhononTransitions}(b)
show the distribution of outgoing (reflected and transmitted) phonon
channels in CNT-12 {[}Fig.~\ref{fig:PhononTransitions}(a){]} and
CNT-24 {[}Fig.~\ref{fig:PhononTransitions}(b){]} as well as the
incoming phonon channel with the wave vector $k_{1}$ in CNT-12 superimposed
on the phonon dispersion spectrum of CNT-12 and CNT-24. The transition
probabilities between the incoming phonon channel at $k_{1}$ and
its main outgoing phonon channels at $\bar{k}_{1}$, $k_{2}$ and
$k_{3}$, which are all doubly degenerate, are calculated from the
matrix elements of $\bar{\boldsymbol{r}}_{\text{LL}}$ and $\bar{\boldsymbol{t}}_{\text{RL}}$
and indicated in Figs.~\ref{fig:PhononTransitions}(a) and (b). The
dominant transition probabilities {[}$P(k_{1}\rightarrow\bar{k}_{1})$,
$P(k_{1}\rightarrow k_{2})$ and $P(k_{1}\rightarrow k_{3})${]} add
up to nearly unity once the two-fold degeneracy of the final phonon
states is taken into account. 

We find that the transmission of the incoming phonon mode at $k_{1}$,
which has a transmission coefficient of $\Xi_{\text{L},1}=0.642$,
is dominated by forward scattering transitions ($k_{1}\rightarrow k_{3}$)
to the outgoing phonon channels at $k_{3}$, with the transition probability
given by $P(k_{1}\rightarrow k_{3})=0.307$ or nearly half of the
transmission coefficient, because the phonon subbands for $k_{3}$
are the CNT-24 image of the phonon subbands for $k_{1}$ as shown
in Figs.~\ref{fig:PhononTransitions}(a) and (b), indicating that
angular symmetry and polarization considerations play an important
role in forward scattering. The phonon reflection processes is dominated
by backward scattering to the phonon channels at $\bar{k}_{1}$ and
$k_{2}$. Unusually, the $k_{1}\rightarrow\bar{k}_{1}$ transition,
which corresponds to an \emph{intra}-subband process, has a slightly
lower probability than the $k_{1}\rightarrow k_{2}$ transition, an
\emph{inter}-subband process, suggesting that transitions between
these two phonon subbands, indicated by bold dashed and dotted lines
in  panels (a) and (b), are favored in backward scattering. 

\subsubsection{Incoming phonon channel at $\bar{k}_{3}$ in CNT-24}

\begin{figure*}

\includegraphics[width=14cm]{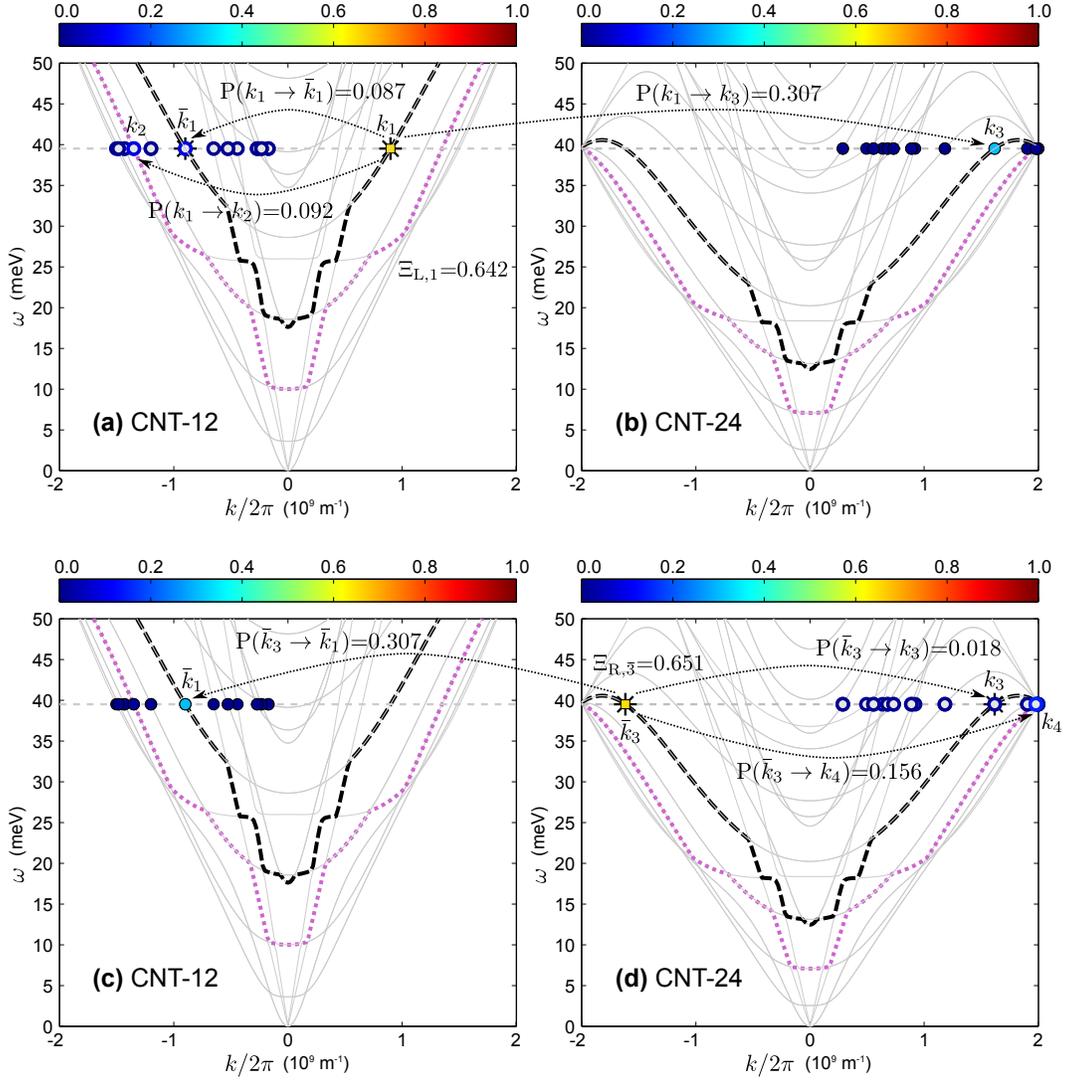}

\caption{Plot of leftward-going phonon channels (hollow circles) in \textbf{(a)}
CNT-12 and rightward-going phonon channels (solid circles) in \textbf{(b)}
CNT-24 for the incoming phonon channel (square symbol) at $\omega=39.5$
meV and $k_{1}=5.65\times10^{9}$ m$^{-1}$ in CNT-12, superimposed
on the phonon dispersion curves of CNT-12 and CNT-24. The transition
probability for each outgoing phonon channel from the incoming phonon
at $k_{1}$ is between 0 and 1, and indicated in color. The dominant
scattering processes, corresponding to (i) the $k_{1}\rightarrow\bar{k}_{1}$
intra-subband reflection, (ii) the $k_{1}\rightarrow k_{2}$ inter-subband
reflection, and (iii) the $k_{1}\rightarrow k_{3}$ transmission,
are drawn with dotted arrows with the transition probabilities explicitly
given. The outgoing phonon channels in \textbf{(c)} CNT-12 and \textbf{(d)}
CNT-24 for the incoming phonon channel at $\omega=39.5$ meV and $\bar{k}_{3}$
in CNT-24 are also shown. The main outgoing phonon channels for the
incoming phonon at $\bar{k}_{3}$ are at $\bar{k}_{1}$, $k_{3}$
and $k_{4}$ and also indicated by dotted arrows with the transition
probabilities given. To guide the eye, the subbands for $k_{1}$,
$\bar{k}_{1}$, $k_{3}$ and $\bar{k}_{3}$ are indicated in bold
dashed lines while the subbands for $k_{2}$ and $k_{4}$ are indicated
in bold magenta dotted lines. }

\label{fig:PhononTransitions}
\end{figure*}

Given the dominant transition between $k_{1}$ in CNT-12 and $k_{3}$
in CNT-24, it would be interesting to study the scattering processes
associated with the incoming phonon channel at $\bar{k}_{3}$ in CNT-24.
As before, the transition probabilities are computed from the matrix
elements of $\bar{\boldsymbol{r}}_{\text{RR}}$ and $\bar{\boldsymbol{t}}_{\text{LR}}$,
and shown in Figs.~\ref{fig:PhononTransitions}(c) and (d). We find
that the transmission of the mode at $\bar{k}_{3}$, which has a transmission
coefficient of $\Xi_{\text{R},\bar{3}}=0.651$, is dominated by the
$\bar{k}_{3}\rightarrow\bar{k}_{1}$ process which has the transition
probability of $P(\bar{k}_{3}\rightarrow\bar{k}_{1})=0.307$, numerically
equal to $P(k_{1}\rightarrow k_{3})$ as expected, because the $\bar{k}_{3}\rightarrow\bar{k}_{1}$
transition is the time reversal of the $k_{1}\rightarrow k_{3}$ transition
in Figs.~\ref{fig:PhononTransitions}(a) and (b). Also, the main
reflected outgoing phonon channels in CNT-24 are at $k_{3}$ and $k_{4}$.
Like in the previous simulation, the $\bar{k}_{3}\rightarrow k_{4}$
transition, an inter-subband process, plays a greater role in phonon
reflection than the $\bar{k}_{3}\rightarrow k_{3}$ transition, an
intra-subband process, but also to a substantially greater extent
since $P(\bar{k}_{3}\rightarrow k_{4})\gg P(\bar{k}_{3}\rightarrow k_{3})$,
highlighting the role of polarization in phonon scattering. The $\bar{k}_{3}\rightarrow k_{4}$
inter-subband transition is favored because the subband for $k_{4}$
is the CNT-24 image of the subband for $k_{2}$ in Fig.~\ref{fig:PhononTransitions}(a).

\section{Example with zigzag and armchair graphene edge}

To illustrate the utility of our method for studying boundary scattering,
we apply the $S$-matrix method to  investigate the effects of edge
orientation and structure on phonon scattering in graphene. Unlike
the previous example of the CNT junction, there is no phonon transmission
as we are dealing with pure phonon reflection in which every incoming
phonon is backscattered elastically into a range of outgoing phonon
channels. The phonon scattering specularity, important for understanding
phonon transport in graphene nanoribbons,\citep{JHu:NL09_Thermal,MBae:NatCommun13_Ballistic,AMajee:PRB16_Length}
can be obtained from the distribution of the transition probabilities. 

\begin{figure}
\includegraphics[width=8cm]{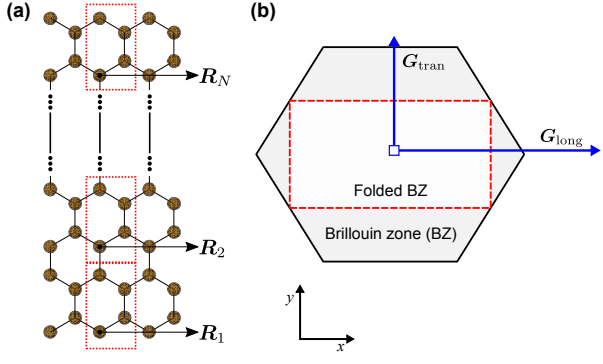}

\caption{\textbf{(a)} Schematic of the bulk graphene slice (bounded by dotted
lines) for the armchair edge scattering simulation. Each slice is
partitioned in the transverse ($y$) direction into 4-atom unit cells.
\textbf{(b)} The 4-atom unit cell is twice as large as the 2-atom
primitive unit cell, resulting in a smaller folded Brillouin zone
(bounded by dashed lines) with half the area of the standard Brillouin
zone (bounded by solid lines). The longitudinal and transverse reciprocal
lattice vectors for the folded BZ are given by $\boldsymbol{G}_{\text{long}}$
and $\boldsymbol{G}_{\text{tran}}$, respectively. }
\label{fig:SliceSchematics}
\end{figure}

In addition, because the system is a two-dimensional one in which
we partition the lattice into unit cells larger than the usual primitive
unit cell, two additional intermediate procedures are needed in the
application of our $S$-matrix method to graphene. The first procedure
deals with the periodic boundary conditions in the transverse direction
which affect the structure of the matrices $\boldsymbol{H}_{\text{L}}^{00}$
and $\boldsymbol{H}_{\text{L}}^{01}$ associated with the bulk lead
and permit us to decompose them into their Fourier-component submatrices,
facilitating the efficient computation of the surface and bulk Green's
functions. This Fourier decomposition requires us to partition the
rectangular slices in Fig.~\ref{fig:SystemSchematic} into unit cells
in the transverse direction {[}Fig.~\ref{fig:SliceSchematics}(a){]}
and index the incoming and outgoing phonon channels with wave vectors
associated with phonon modes in the `folded' Brillouin zone {[}Fig.~\ref{fig:SliceSchematics}(b){]}
which follows from the transverse partitioning of the rectangular
slices in Fig.~\ref{fig:SystemSchematic}. The second procedure deals
with the mapping of the phonon modes in the `folded' Brillouin zone
to the bulk phonon eigenmodes in the standard `unfolded' Brillouin
zone associated with the symmetry of the primitive unit cell in graphene.
Although this step is not strictly necessary, the use of the zone-unfolding
technique, as described by Boykin and Klimeck,~\citep{TBoykin:PRB05_Practical,TBoykin:PhysicaE09_Non}
improves the clarity of the scattering results by presenting their
analysis in more familiar terms. 

\subsection{Calculation details}

Like in the previous example, we construct the bulk graphene monolayer
and optimize its structure in GULP~\citep{JGale:MolSim03_gulp} using
the same Tersoff potential parameters.~\citep{LLindsay:PRB10_Optimized}.
We assume that the graphene edge is terminated on the right and its
bulk extends infinitely to the left. Thus, unlike the schematic shown
in Fig.~(\ref{fig:SystemSchematic}), we need only to consider the
force-constant matrices $\boldsymbol{H}_{\text{L}}^{00}$ and $\boldsymbol{H}_{\text{L}}^{01}$
to describe the left bulk and $\boldsymbol{H}_{\text{C}}$ and $\boldsymbol{H}_{\text{CL}}$
to describe the graphene edge and its coupling to the left bulk. The
force-constant matrices $\boldsymbol{H}_{\text{R}}^{00}$, $\boldsymbol{H}_{\text{R}}^{01}$
and $\boldsymbol{H}_{\text{CR}}$ in Eq.~(\ref{eq:SystemForceConstantMatrix})
are not needed in this study and their matrix elements are set to
zero. 

The force-constant matrices for the bulk slices ($\boldsymbol{H}_{\text{L}}^{00}$
and $\boldsymbol{H}_{\text{L}}^{01}$) are computed in GULP. For the
armchair and zigzag edge structures, the slices in the leads each
have $4N$ atoms. We take advantage of the periodicity in the transverse
direction to partition the slice into $N$ 4-atom unit cells, as shown
in Fig.~\ref{fig:SliceSchematics}(a), at the real lattice points
$\boldsymbol{R}_{1},\ldots,\boldsymbol{R}_{N}$ where $\boldsymbol{R}_{p}=(p-1)\boldsymbol{T}$
and $\boldsymbol{T}$ is the lattice vector characterizing the transverse
periodicity. The $12\times12$ force-constant submatrix corresponding
to the coupling between the unit cells at $\boldsymbol{R}_{p}$ and
$\boldsymbol{R}_{q}$ within the same slice is denoted as $\mathbf{H}_{\text{L}}^{00}(\boldsymbol{R}_{p},\boldsymbol{R}_{q})$
while the $12\times12$ force-constant submatrix corresponding to
the coupling between the unit cell at $\boldsymbol{R}_{p}$ in the
slice and the unit cell at $\boldsymbol{R}_{q}$ in the slice on the
right (left) is denoted by $\mathbf{H}_{\text{L}}^{01}(\boldsymbol{R}_{p},\boldsymbol{R}_{q})$
{[}$\mathbf{H}_{\text{L}}^{10}(\boldsymbol{R}_{p},\boldsymbol{R}_{q})${]}. 

The transverse translational symmetry implies that the force-constant
submatrices depend only on the relative displacement between the unit
cells in the transverse direction, i.e. 
\begin{equation}
\mathbf{H}_{\text{L}}^{lm}(\boldsymbol{R}_{p},\boldsymbol{R}_{q})=\mathbf{H}_{\text{L}}^{lm}(\boldsymbol{R}_{p}-\boldsymbol{R}_{q})\label{eq:TranslationalInvarianceCondition}
\end{equation}
for $l=0,1$ and $m=(l-1)\mod2$. For a slice with $N$ transverse
unit cells, the submatrices make up the $12N\times12N$ matrix associated
with the entire slice, \begin{widetext}
\begin{align}
\boldsymbol{H}_{\text{L}}^{lm}=\left(\begin{array}{cccc}
\mathbf{H}_{\text{L}}^{lm}(\boldsymbol{R}_{1},\boldsymbol{R}_{1}) & \mathbf{H}_{\text{L}}^{lm}(\boldsymbol{R}_{1},\boldsymbol{R}_{2}) & \cdots & \mathbf{H}_{\text{L}}^{lm}(\boldsymbol{R}_{1},\boldsymbol{R}_{N})\\
\mathbf{H}_{\text{L}}^{lm}(\boldsymbol{R}_{2},\boldsymbol{R}_{1}) & \mathbf{H}_{\text{L}}^{lm}(\boldsymbol{R}_{2},\boldsymbol{R}_{2}) & \cdots & \mathbf{H}_{\text{L}}^{lm}(\boldsymbol{R}_{2},\boldsymbol{R}_{N})\\
\vdots & \vdots & \ddots & \vdots\\
\mathbf{H}_{\text{L}}^{lm}(\boldsymbol{R}_{N},\boldsymbol{R}_{1}) & \mathbf{H}_{\text{L}}^{lm}(\boldsymbol{R}_{N},\boldsymbol{R}_{2}) & \cdots & \mathbf{H}_{\text{L}}^{lm}(\boldsymbol{R}_{N},\boldsymbol{R}_{N})
\end{array}\right)\ .\label{eq:HL_definition}
\end{align}
It follows from Eq.~(\ref{eq:HL_definition}) that $\mathbf{H}_{\text{L}}^{ml}(\boldsymbol{R}_{q},\boldsymbol{R}_{p})=\mathbf{H}_{\text{L}}^{lm}(\boldsymbol{R}_{p},\boldsymbol{R}_{q})^{\dagger}$.
In addition, Eq.~(\ref{eq:TranslationalInvarianceCondition}) and
the transverse periodic boundary conditions imply that we can write
Eq.~(\ref{eq:HL_definition}) as
\begin{equation}
\boldsymbol{H}_{\text{L}}^{lm}=\left(\begin{array}{cccc}
\mathbf{H}_{\text{L}}^{lm}(0) & \mathbf{H}_{\text{L}}^{lm}(-\boldsymbol{T}) & \cdots & \mathbf{H}_{\text{L}}^{lm}(-(N-1)\boldsymbol{T})\\
\mathbf{H}_{\text{L}}^{lm}(-(N-1)\boldsymbol{T}) & \mathbf{H}_{\text{L}}^{lm}(0) & \cdots & \mathbf{H}_{\text{L}}^{lm}(-(N-2)\boldsymbol{T})\\
\vdots & \vdots & \ddots & \vdots\\
\mathbf{H}_{\text{L}}^{lm}(-\boldsymbol{T}) & \mathbf{H}_{\text{L}}^{lm}(-2\boldsymbol{T}) & \cdots & \mathbf{H}_{\text{L}}^{lm}(0)
\end{array}\right)\ ,\label{eq:ToeplitzFormHL}
\end{equation}
which has the form of a block-circulant matrix.~\citep{DeMazancout:IEEETrans83_Inverse} 

\subsubsection{Working with transverse Fourier components }

Although it seems natural to use Eq.~(\ref{eq:AllRetardedSurfaceGF})
directly to determine the surface Green's function, it is numerically
more efficient to exploit the block-circulant matrix structure of
Eq.~(\ref{eq:ToeplitzFormHL}) by employing a discrete Fourier-transform
approach like in Ref.~\citep{DeMazancout:IEEETrans83_Inverse} which
also yields a set of indices $\boldsymbol{Q}_{n}$, where $n=0,\ldots,N-1$,
associated with the periodicity in the transverse direction. The matrix
$\boldsymbol{H}_{\text{L}}^{lm}$ in Eq.~(\ref{eq:ToeplitzFormHL})
can be transformed into the block-diagonal form $\tilde{\boldsymbol{H}}_{\text{L}}^{lm}$,
via the expression 
\begin{equation}
\boldsymbol{H}_{\text{L}}^{lm}=\boldsymbol{P}\tilde{\boldsymbol{H}}_{\text{L}}^{lm}\boldsymbol{P}^{-1}\label{eq:CirculantMatrixFormHL}
\end{equation}
where 
\begin{equation}
\boldsymbol{P}=\frac{1}{\sqrt{N}}\left(\begin{array}{cccc}
\tilde{\mathbf{I}}_{\text{L}}e^{i\boldsymbol{Q}_{1}\cdot\boldsymbol{R}_{1}} & \tilde{\mathbf{I}}_{\text{L}}e^{i\boldsymbol{Q}_{2}\cdot\boldsymbol{R}_{1}} & \cdots & \tilde{\mathbf{I}}_{\text{L}}e^{i\boldsymbol{Q}_{N}\cdot\boldsymbol{R}_{1}}\\
\tilde{\mathbf{I}}_{\text{L}}e^{i\boldsymbol{Q}_{1}\cdot\boldsymbol{R}_{2}} & \tilde{\mathbf{I}}_{\text{L}}e^{i\boldsymbol{Q}_{2}\cdot\boldsymbol{R}_{2}} & \cdots & \tilde{\mathbf{I}}_{\text{L}}e^{i\boldsymbol{Q}_{N}\cdot\boldsymbol{R}_{2}}\\
\vdots & \vdots & \ddots & \vdots\\
\tilde{\mathbf{I}}_{\text{L}}e^{i\boldsymbol{Q}_{1}\cdot\boldsymbol{R}_{N}} & \tilde{\mathbf{I}}_{\text{L}}e^{i\boldsymbol{Q}_{2}\cdot\boldsymbol{R}_{N}} & \cdots & \tilde{\mathbf{I}}_{\text{L}}e^{i\boldsymbol{Q}_{N}\cdot\boldsymbol{R}_{N}}
\end{array}\right)\label{eq:UnitaryTransformMatrix}
\end{equation}
is the special unitary matrix used for the basis transformation, $\tilde{\mathbf{I}}_{\text{L}}$
is the $12\times12$ identity submatrix, and $\tilde{\boldsymbol{H}}_{\text{L}}^{lm}$
is
\begin{equation}
\tilde{\boldsymbol{H}}_{\text{L}}^{lm}=\left(\begin{array}{cccc}
\tilde{\mathbf{H}}_{\text{L}}^{lm}(\boldsymbol{Q}_{1})\\
 & \tilde{\mathbf{H}}_{\text{L}}^{lm}(\boldsymbol{Q}_{2})\\
 &  & \ddots\\
 &  &  & \tilde{\mathbf{H}}_{\text{L}}^{lm}(\boldsymbol{Q}_{N})
\end{array}\right)\ .\label{eq:BlockDiagonalHL}
\end{equation}
Each diagonal submatrix in Eq.~(\ref{eq:BlockDiagonalHL}) is the
discrete Fourier transform of $\mathbf{H}_{\text{L}}^{lm}(\boldsymbol{R}_{p},\boldsymbol{R}_{q})$,
i.e.
\begin{equation}
\tilde{\mathbf{H}}_{\text{L}}^{lm}(\boldsymbol{Q}_{n})=\sum_{q=0}^{N-1}\mathbf{H}_{\text{L}}^{lm}(\boldsymbol{R}_{p},\boldsymbol{R}_{p+q})e^{-i\boldsymbol{Q}_{n}\cdot(\boldsymbol{R}_{p}-\boldsymbol{R}_{p+q})}\label{eq:FourierHSubmatrices}
\end{equation}
where $l=0,1$ and $m=(l-1)\mod2$, and represents a transverse Fourier
component corresponding to the transverse wave vector $\boldsymbol{Q}_{n}=\frac{n}{N}\boldsymbol{G}_{\text{tran}}$,
where $n=0,\ldots,N-1$, and $\boldsymbol{G}_{\text{tran}}$ is the
transverse reciprocal lattice vector satisfying $\boldsymbol{G}_{\text{tran}}\cdot\boldsymbol{T}=2\pi$.
It can also be shown that $\tilde{\mathbf{H}}_{\text{L}}^{ml}(\boldsymbol{Q}_{n})=[\tilde{\mathbf{H}}_{\text{L}}^{lm}(\boldsymbol{Q}_{n})]^{\dagger}$. 

The block-diagonal form of Eq.~(\ref{eq:BlockDiagonalHL}) allows
us to treat each Fourier component as an effectively independent subsystem
and determine piecewise the essential matrix variables such as the
surface Green's functions from the force-constant submatrices $\tilde{\mathbf{H}}_{\text{L}}^{00}(\boldsymbol{Q}_{n})$
and $\tilde{\mathbf{H}}_{\text{L}}^{01}(\boldsymbol{Q}_{n})$, using
the methodology described in Sec.~\ref{sec:Method}. In the following
discussions, we use the $\boldsymbol{B}_{\text{L},\pm}^{\text{adv/ret}}$
as a shorthand notation to refer to the four related matrices $\boldsymbol{B}_{\text{L},+}^{\text{ret}}$,
$\boldsymbol{B}_{\text{L},-}^{\text{ret}}$, $\boldsymbol{B}_{\text{L},+}^{\text{adv}}$
and $\boldsymbol{B}_{\text{L},-}^{\text{adv}}$ where $\boldsymbol{B}$
is any matrix function (e.g. the surface Green's function $\boldsymbol{g}$). 

In the same manner, the surface Green's function can be block-diagonalized
with the same $\boldsymbol{P}$ in Eq.~(\ref{eq:CirculantMatrixFormHL}),
i.e.,
\begin{equation}
\boldsymbol{g}_{\text{L},\pm}^{\text{adv/ret}}=\boldsymbol{P}\tilde{\boldsymbol{g}}_{\text{L},\pm}^{\text{adv/ret}}\boldsymbol{P}^{-1}\label{eq:CirculantFormSurfaceGreensFunction}
\end{equation}
where $\tilde{\boldsymbol{g}}_{\text{L},\pm}^{\text{adv/ret}}$ is
a block-diagonal matrix like $\tilde{\boldsymbol{H}}_{\text{L}}^{lm}$
in Eq.~(\ref{eq:BlockDiagonalHL}) and has the block-diagonal $12\times12$
submatrices $\tilde{\mathbf{g}}_{\text{L},\pm}^{\text{adv/ret}}(\boldsymbol{Q}_{n})$
for $n=1,\ldots,N$, with \begin{subequations}
\begin{equation}
\tilde{\mathbf{g}}_{\text{L},-}^{\text{ret}}(\boldsymbol{Q}_{n})=[(\omega^{2}+i\eta)\tilde{\mathbf{I}}_{\text{L}}-\mathbf{\tilde{H}}_{\text{L}}^{00}(\boldsymbol{Q}_{n})-\mathbf{\tilde{H}}_{\text{L}}^{10}(\boldsymbol{Q}_{n})\tilde{\mathbf{g}}_{\text{L},-}^{\text{ret}}(\boldsymbol{Q}_{n})\mathbf{\tilde{H}}_{\text{L}}^{01}(\boldsymbol{Q}_{n})]^{-1}
\end{equation}
\begin{equation}
\tilde{\mathbf{g}}_{\text{L},+}^{\text{ret}}(\boldsymbol{Q}_{n})=[(\omega^{2}+i\eta)\tilde{\mathbf{I}}_{\text{L}}-\mathbf{\tilde{H}}_{\text{L}}^{00}(\boldsymbol{Q}_{n})-\mathbf{\tilde{H}}_{\text{L}}^{01}(\boldsymbol{Q}_{n})\tilde{\mathbf{g}}_{\text{L},+}^{\text{ret}}(\boldsymbol{Q}_{n})\mathbf{\tilde{H}}_{\text{L}}^{10}(\boldsymbol{Q}_{n})]^{-1}
\end{equation}
\label{eq:FourierTransformRetSurfaceGF}\end{subequations} like in
Eq.~(\ref{eq:AllRetardedSurfaceGF}) and $\tilde{\mathbf{g}}_{\text{L},\pm}^{\text{adv}}(\boldsymbol{Q}_{n})=\tilde{\mathbf{g}}_{\text{L},\pm}^{\text{ret}}(\boldsymbol{Q}_{n})^{\dagger}$. 

Similarly, we have the block-diagonal Bloch matrices $\tilde{\boldsymbol{F}}_{\text{L},\pm}^{\text{adv/ret}}$
with the diagonal submatrices $\tilde{\mathbf{F}}_{\text{L},\pm}^{\text{adv/ret}}(\boldsymbol{Q}_{n})$
given by $\tilde{\mathbf{F}}_{\text{L},+}^{\text{adv/ret}}(\boldsymbol{Q}_{n})=\tilde{\mathbf{g}}_{\text{L},+}^{\text{adv/ret}}(\boldsymbol{Q}_{n})\mathbf{\tilde{H}}_{\text{L}}^{10}(\boldsymbol{Q}_{n})$
and $\tilde{\mathbf{F}}_{\text{L},-}^{\text{adv/ret}}(\boldsymbol{Q}_{n})=\tilde{\mathbf{g}}_{\text{L},-}^{\text{adv/ret}}(\boldsymbol{Q}_{n})\mathbf{\tilde{H}}_{\text{L}}^{01}(\boldsymbol{Q}_{n})$
from Eq.~(\ref{eq:BlochMatrices}). The bulk eigenmode submatrices
$\tilde{\mathbf{U}}_{\text{L},\pm}^{\text{adv/ret}}(\boldsymbol{Q}_{n})$
are determined from Eq.~(\ref{eq:BlochMatrixEigenmodes}), i.e.,
$\tilde{\mathbf{F}}_{\text{L},+}^{\text{adv/ret}}(\boldsymbol{Q}_{n})\tilde{\mathbf{U}}_{\text{L},+}^{\text{adv/ret}}(\boldsymbol{Q}_{n})=\tilde{\mathbf{U}}_{\text{L},+}^{\text{adv/ret}}(\boldsymbol{Q}_{n})\tilde{\boldsymbol{\Lambda}}_{\text{L},+}^{\text{adv/ret}}(\boldsymbol{Q}_{n})$
and $\tilde{\mathbf{F}}_{\text{L},-}^{\text{adv/ret}}(\boldsymbol{Q}_{n})^{-1}\tilde{\mathbf{U}}_{\text{L},-}^{\text{adv/ret}}(\boldsymbol{Q}_{n})=\tilde{\mathbf{U}}_{\text{L},-}^{\text{adv/ret}}(\boldsymbol{Q}_{n})\tilde{\boldsymbol{\Lambda}}_{\text{L},-}^{\text{adv/ret}}(\boldsymbol{Q}_{n})^{-1}$.
As in Eq.~(\ref{eq:BlochMatrixEigenmodes}), the matrices $\tilde{\boldsymbol{\Lambda}}_{\text{L},\pm}^{\text{adv/ret}}(\boldsymbol{Q}_{n})$
have only diagonal elements containing the eigenvalues of $\tilde{\mathbf{F}}_{\text{L},\pm}^{\text{adv/ret}}$
and make up the block-diagonal submatrices in 
\begin{equation}
\tilde{\boldsymbol{\Lambda}}_{\text{L},\pm}^{\text{adv/ret}}=\left(\begin{array}{cccc}
\tilde{\boldsymbol{\Lambda}}_{\text{L},\pm}^{\text{adv/ret}}(\boldsymbol{Q}_{1})\\
 & \tilde{\boldsymbol{\Lambda}}_{\text{L},\pm}^{\text{adv/ret}}(\boldsymbol{Q}_{2})\\
 &  & \ddots\\
 &  &  & \tilde{\boldsymbol{\Lambda}}_{\text{L},\pm}^{\text{adv/ret}}(\boldsymbol{Q}_{N})
\end{array}\right)
\end{equation}
which is a purely diagonal $12N\times12N$ matrix. The Bloch eigenmode
matrices have the form
\begin{equation}
\tilde{\mathbf{U}}_{\text{L},\pm}^{\text{adv/ret}}(\boldsymbol{Q}_{n})=\left(\begin{array}{ccc}
\tilde{\mathbf{u}}_{\text{L},\pm}^{\text{adv/ret}}(\boldsymbol{Q}_{n},k_{n,1}), & \ldots, & \tilde{\mathbf{u}}_{\text{L},\pm}^{\text{adv/ret}}(\boldsymbol{Q}_{n},k_{n,12})\end{array}\right)\label{eq:BlochEigenmodeMatrixColumnForm}
\end{equation}
where $\tilde{\mathbf{u}}_{\text{L},\pm}^{\text{adv/ret}}(\boldsymbol{Q}_{n},k_{n,m})$
is the $12\times1$ column eigenvector for the transverse wave vector
$\boldsymbol{Q}_{n}$ and the longitudinal wave vector $k_{n,m}$
for $m=1,\ldots,12$. The corresponding eigenvelocity submatrices
$\tilde{\mathbf{V}}_{\text{L},\pm}^{\text{adv/ret}}(\boldsymbol{Q}_{n})$
can be found using Eqs.~(\ref{eq:RightGoingVelocityMatrix}) and
(\ref{eq:LeftGoingVelocityMatrix}), and have the form
\[
\tilde{\mathbf{V}}_{\text{L},\pm}^{\text{adv/ret}}(\boldsymbol{Q}_{n})=\left(\begin{array}{ccc}
v_{\text{L},\pm}^{\text{adv/ret}}(\boldsymbol{Q}_{n},k_{n,1}) & \cdots & 0\\
\vdots & \ddots & \vdots\\
0 & \cdots & v_{\text{L},\pm}^{\text{adv/ret}}(\boldsymbol{Q}_{n},k_{n,12})
\end{array}\right)
\]
where is $v_{\text{L},\pm}^{\text{adv/ret}}(\boldsymbol{Q}_{n},k_{n,m})$
is the corresponding longitudinal group velocity for the eigenmode
$\tilde{\mathbf{u}}_{\text{L},\pm}^{\text{adv/ret}}(\boldsymbol{Q}_{n},k_{n,m})$.

\subsubsection{Real space matrix variables}

To recover the real-space surface Green's function matrix $\boldsymbol{g}_{\text{L},\pm}^{\text{adv/ret}}$,
we apply the transformation $\boldsymbol{g}_{\text{L},\pm}^{\text{adv/ret}}=\boldsymbol{P}\tilde{\boldsymbol{g}}_{\text{L},\pm}^{\text{adv/ret}}\boldsymbol{P}^{-1}$
like in Eq.~(\ref{eq:CirculantFormSurfaceGreensFunction}) and obtain
\begin{equation}
\boldsymbol{g}_{\text{L},\pm}^{\text{adv/ret}}=\left(\begin{array}{cccc}
\mathbf{g}_{\text{L},\pm}^{\text{adv/ret}}(\boldsymbol{R}_{1},\boldsymbol{R}_{1}) & \mathbf{g}_{\text{L},\pm}^{\text{adv/ret}}(\boldsymbol{R}_{1},\boldsymbol{R}_{2}) & \cdots & \mathbf{g}_{\text{L},\pm}^{\text{adv/ret}}(\boldsymbol{R}_{1},\boldsymbol{R}_{N})\\
\mathbf{g}_{\text{L},\pm}^{\text{adv/ret}}(\boldsymbol{R}_{2},\boldsymbol{R}_{1}) & \mathbf{g}_{\text{L},\pm}^{\text{adv/ret}}(\boldsymbol{R}_{2},\boldsymbol{R}_{2}) & \cdots & \mathbf{g}_{\text{L},\pm}^{\text{adv/ret}}(\boldsymbol{R}_{2},\boldsymbol{R}_{N})\\
\vdots & \vdots & \ddots & \vdots\\
\mathbf{g}_{\text{L},\pm}^{\text{adv/ret}}(\boldsymbol{R}_{N},\boldsymbol{R}_{1}) & \mathbf{g}_{\text{L},\pm}^{\text{adv/ret}}(\boldsymbol{R}_{N},\boldsymbol{R}_{2}) & \cdots & \mathbf{g}_{\text{L},\pm}^{\text{adv/ret}}(\boldsymbol{R}_{N},\boldsymbol{R}_{N})
\end{array}\right)\ .\label{eq:RealSpaceSurfaceGF}
\end{equation}
Similarly, the real-space Bloch matrix from Eq.~(\ref{eq:BlochMatrices})
can be obtained via the expression $\boldsymbol{F}_{\text{L}}^{\text{adv/ret}}(\pm)=\boldsymbol{P}\tilde{\boldsymbol{F}}_{\text{L},\pm}^{\text{adv/ret}}\boldsymbol{P}^{-1}$.
Given that the real-space Bloch matrix must satisfy the conditions
\begin{equation}
\boldsymbol{F}_{\text{L}}^{\text{adv/ret}}(\pm)^{\pm1}\boldsymbol{U}_{\text{L}}^{\text{adv/ret}}(\pm)=\boldsymbol{U}_{\text{L}}^{\text{adv/ret}}(\pm)\boldsymbol{\Lambda}_{\text{L}}^{\text{adv/ret}}(\pm)^{\pm}\label{eq:RealSpaceBlochMatrix}
\end{equation}
where $\boldsymbol{\Lambda}_{\text{L}}^{\text{adv/ret}}(\pm)$ is
also a purely diagonal matrix like $\tilde{\boldsymbol{\Lambda}}_{\text{L},\pm}^{\text{adv/ret}}$
with the eigenvalues of $\boldsymbol{F}_{\text{L}}^{\text{adv/ret}}(\pm)$
along its diagonal. Equation~(\ref{eq:RealSpaceBlochMatrix}) implies
that $\boldsymbol{\Lambda}_{\text{L}}^{\text{adv/ret}}(\pm)=\tilde{\boldsymbol{\Lambda}}_{\text{L},\pm}^{\text{adv/ret}}$
and we can write the real-space bulk eigenmode matrix as 
\begin{equation}
\boldsymbol{U}_{\text{L}}^{\text{adv/ret}}(\pm)=\boldsymbol{P}\tilde{\boldsymbol{U}}_{\text{L},\pm}^{\text{adv/ret}}\ ,\label{eq:RealSpaceEigenmodes}
\end{equation}
giving us 

\begin{align}
\boldsymbol{U}_{\text{L}}^{\text{adv/ret}}(\pm) & =\frac{1}{\sqrt{N}}\left(\begin{array}{cccc}
\tilde{\mathbf{U}}_{\text{L},\pm}^{\text{adv/ret}}(\boldsymbol{Q}_{1})e^{i\boldsymbol{Q}_{1}\cdot\boldsymbol{R}_{1}} & \tilde{\mathbf{U}}_{\text{L},\pm}^{\text{adv/ret}}(\boldsymbol{Q}_{2})e^{i\boldsymbol{Q}_{2}\cdot\boldsymbol{R}_{1}} & \cdots & \tilde{\mathbf{U}}_{\text{L},\pm}^{\text{adv/ret}}(\boldsymbol{Q}_{N})e^{i\boldsymbol{Q}_{N}\cdot\boldsymbol{R}_{1}}\\
\tilde{\mathbf{U}}_{\text{L},\pm}^{\text{adv/ret}}(\boldsymbol{Q}_{1})e^{i\boldsymbol{Q}_{1}\cdot\boldsymbol{R}_{2}} & \tilde{\mathbf{U}}_{\text{L},\pm}^{\text{adv/ret}}(\boldsymbol{Q}_{2})e^{i\boldsymbol{Q}_{2}\cdot\boldsymbol{R}_{2}} & \cdots & \tilde{\mathbf{U}}_{\text{L},\pm}^{\text{adv/ret}}(\boldsymbol{Q}_{N})e^{i\boldsymbol{Q}_{N}\cdot\boldsymbol{R}_{2}}\\
\vdots & \vdots & \ddots & \vdots\\
\tilde{\mathbf{U}}_{\text{L},\pm}^{\text{adv/ret}}(\boldsymbol{Q}_{1})e^{i\boldsymbol{Q}_{1}\cdot\boldsymbol{R}_{N}} & \tilde{\mathbf{U}}_{\text{L},\pm}^{\text{adv/ret}}(\boldsymbol{Q}_{2})e^{i\boldsymbol{Q}_{2}\cdot\boldsymbol{R}_{N}} & \cdots & \tilde{\mathbf{U}}_{\text{L},\pm}^{\text{adv/ret}}(\boldsymbol{Q}_{N})e^{i\boldsymbol{Q}_{N}\cdot\boldsymbol{R}_{N}}
\end{array}\right)\label{eq:RealSpaceEigenmodeMatrix}\\
 & =\left(\begin{array}{c}
\mathbf{u}_{\text{L},\pm}^{\text{adv/ret}}(\boldsymbol{Q}_{1},k_{1,1}),\ldots,\mathbf{\mathbf{u}}_{\text{L},\pm}^{\text{adv/ret}}(\boldsymbol{Q}_{1},k_{1,12}),\ldots,\mathbf{\mathbf{u}}_{\text{L},\pm}^{\text{adv/ret}}(\boldsymbol{Q}_{N},k_{N,1}),\ldots,\mathbf{u}_{\text{L},\pm}^{\text{adv/ret}}(\boldsymbol{Q}_{N},k_{N,12})\end{array}\right)\nonumber 
\end{align}
where the right-hand side of Eq.~(\ref{eq:RealSpaceEigenmodeMatrix})
is a $12N\times12N$ matrix with each column vector corresponding
to an extended or evanescent bulk eigenmode and represented by $\mathbf{u}_{\text{L},\pm}^{\text{adv/ret}}(\boldsymbol{Q}_{n},k_{n,m})$,
where $n=1,\ldots,N$ and $m=1,\ldots,12$. Hence, we have a total
of $12N$ eigenmodes, associated with each is a real or complex longitudinal
wave vector. For each transverse wave vector $\boldsymbol{Q}_{n}$,
we have $12$ longitudinal wave vectors which we enumerate as $k_{n,1}$
to $k_{n,12}$. It also follows from Eqs.~(\ref{eq:RealSpaceBlochMatrix})
and (\ref{eq:RealSpaceEigenmodes}) that the real-space velocity matrix
is $\boldsymbol{V}_{\text{L}}^{\text{adv/ret}}(\pm)=\tilde{\boldsymbol{V}}_{\text{L},\pm}^{\text{adv/ret}}$. 

Given the real-space surface Green's functions in Eq.~(\ref{eq:RealSpaceSurfaceGF}),
we can compute the effective harmonic matrix in Eq.~(\ref{eq:ProjectedForceConstantMatrix})
and the corresponding Green's function $\boldsymbol{G}_{\text{L}}^{\text{ret}}$
from Eq.~(\ref{eq:FiniteGreensFunction}). Using $\boldsymbol{V}_{\text{L}}^{\text{ret/adv}}(-)$
and $\boldsymbol{U}_{\text{L}}^{\text{ret/adv}}(-)$ from Eq.~(\ref{eq:RealSpaceEigenmodeMatrix}),
we compute $\bar{\boldsymbol{r}}_{\text{LL}}$ from Eq.~(\ref{eq:rmatrix_LL})
which gives us the transition amplitudes between the incoming and
outgoing phonon channels.

\end{widetext}

\subsubsection{Brillouin zone unfolding}

In our transverse partitioning scheme, we can associate with each
phonon channel in Eq.~(\ref{eq:RealSpaceEigenmodeMatrix}) a transverse
wave vector $\boldsymbol{Q}_{n}$ and its longitudinal wave vector
$k_{n,m}$. The vector sum of these two wave vectors ($\boldsymbol{k}=\boldsymbol{Q}_{n}+k_{n,m}\hat{\boldsymbol{x}}$
where the longitudinal direction is in the $x$ direction) yields
the locus of the mode ($\boldsymbol{k}$) within the `folded' Brillouin
zone (BZ) as shown in Fig.~\ref{fig:SliceSchematics}(b). This folded
BZ is a consequence of the 4-atom unit supercell used in our $S$-matrix
method, which requires the partitioning of the atomic degrees of freedom
into rectangular slices, and thus has half the reciprocal space area
of the primitive BZ but contains 12 phonon branches compared to 6
phonon branches in the primitive BZ. 

To make sense of our analysis of the transmission coefficients and
individual transition amplitudes, it is necessary to map the scattering
channels to the phonon modes in the \emph{bulk} graphene lattice.
This is done by `unfolding' the 12 phonon branches within the folded
BZ to obtain 6 phonon branches within the larger primitive BZ using
the zone-unfolding technique of Boykin and Klimeck.~\citep{TBoykin:PRB05_Practical,TBoykin:PhysicaE09_Non}
Given our choice of the 4-atom unit supercell, each phonon mode ($\boldsymbol{k}$)
in the folded BZ has two possible image points ($\boldsymbol{k}^{\prime}$)
in the primitive BZ, with one of them satisfying $\boldsymbol{k}^{\prime}=\boldsymbol{k}$
and the other shifted by an integer multiple of $\boldsymbol{G}_{\text{long}}$
and $\boldsymbol{G}_{\text{tran}}$, i.e. $\boldsymbol{k}^{\prime}=\boldsymbol{k}+n_{1}\boldsymbol{G}_{\text{long}}+n_{2}\boldsymbol{G}_{\text{tran}}$,
where $n_{1}$ and $n_{2}$ are whole numbers that depend on $\boldsymbol{k}$.
For notational brevity, we write $\boldsymbol{k}^{\prime}=\boldsymbol{k}+\boldsymbol{G}(\boldsymbol{k})$.
However, only one of the two image points corresponds to the correct
bulk mode, except in the special case where $\boldsymbol{Q}_{n}=-\boldsymbol{G}_{\text{tran}}/2$
and all the phonon modes are two-fold degenerate. 

For completeness, we outline the application of the Boykin-Klimeck
unfolding technique~\citep{TBoykin:PRB05_Practical,TBoykin:PhysicaE09_Non}
to the graphene lattice. We write the $12\times1$ column eigenvector
$\tilde{\mathbf{u}}_{\text{L},\pm}^{\text{adv/ret}}(\boldsymbol{Q}_{n},k_{n,m})$
in Eq.~(\ref{eq:BlochEigenmodeMatrixColumnForm}), after dropping
the superscripts and subscripts for the sake of brevity, as 
\begin{equation}
\tilde{\mathbf{u}}(\boldsymbol{k})=\left(\begin{array}{c}
\tilde{\boldsymbol{\beta}}_{1}(\boldsymbol{k})\\
\tilde{\boldsymbol{\beta}}_{2}(\boldsymbol{k})
\end{array}\right)\ ,\label{eq:Eigenmode_k}
\end{equation}
where, for $n=1,2$, $\tilde{\boldsymbol{\beta}}_{n}(\boldsymbol{k})$
is the $6\times1$ column vector corresponding to $n$-th 2-atom primitive
unit cell of the 4-atom supercell, and $\boldsymbol{\rho}_{n}$ is
its displacement vector within the supercell. From Eq.~(\ref{eq:Eigenmode_k}),
we define the $12\times1$ column vector
\[
\tilde{\boldsymbol{B}}(\boldsymbol{k})=\left(\begin{array}{c}
\tilde{\boldsymbol{\beta}}_{1}(\boldsymbol{k})e^{-i\boldsymbol{k}\cdot\boldsymbol{\rho}_{1}}\\
\tilde{\boldsymbol{\beta}}_{2}(\boldsymbol{k})e^{-i\boldsymbol{k}\cdot\boldsymbol{\rho}_{2}}
\end{array}\right)
\]
and the $12\times12$ matrix
\[
\boldsymbol{W}(\boldsymbol{k})=\frac{1}{\sqrt{2}}\left(\begin{array}{cc}
\tilde{\boldsymbol{I}} & \tilde{\boldsymbol{I}}e^{i\boldsymbol{G}(\boldsymbol{k})\cdot\boldsymbol{\rho}_{1}}\\
\tilde{\boldsymbol{I}} & \tilde{\boldsymbol{I}}e^{i\boldsymbol{G}(\boldsymbol{k})\cdot\boldsymbol{\rho}_{2}}
\end{array}\right)\ ,
\]
where $\tilde{\boldsymbol{I}}$ is the $6\times6$ identity matrix.
The $12\times1$ column vector $\tilde{\boldsymbol{C}}(\boldsymbol{k})$
containing the unfolded modes is given by~\citep{TBoykin:PRB05_Practical,TBoykin:PhysicaE09_Non}
\begin{equation}
\tilde{\boldsymbol{C}}(\boldsymbol{k})=\left(\begin{array}{c}
\tilde{\mathbf{c}}_{\boldsymbol{k}}\\
\tilde{\mathbf{c}}_{\boldsymbol{k}+\boldsymbol{G}(\boldsymbol{k})}
\end{array}\right)=\boldsymbol{W}(\boldsymbol{k})^{-1}\tilde{\boldsymbol{B}}(\boldsymbol{k})\label{eq:PossibleUnfoldedModes}
\end{equation}
where $\tilde{\mathbf{c}}_{\boldsymbol{k}}$ and $\tilde{\mathbf{c}}_{\boldsymbol{k}+\boldsymbol{G}(\boldsymbol{k})}$
are the $6\times1$ column vectors corresponding to the the possible
unfolded eigenmodes at $\boldsymbol{k}^{\prime}=\boldsymbol{k}$ and
$\boldsymbol{k}^{\prime}=\boldsymbol{k}+\boldsymbol{G}(\boldsymbol{k})$,
respectively. If the folded mode in Eq.~(\ref{eq:Eigenmode_k}) is
not degenerate, then only one of the two possible unfolded eigenmodes
in Eq.~(\ref{eq:PossibleUnfoldedModes}) is correct and the correct
unfolded wave vector can be identified through elimination as the
incorrect eigenmode is zero in all its components. Using Eq.~(\ref{eq:PossibleUnfoldedModes})
as an example, if $|\tilde{\mathbf{c}}_{\boldsymbol{k}}|=0$, then
the correct unfolded wave vector is $\boldsymbol{k}^{\prime}=\boldsymbol{k}+\boldsymbol{G}(\boldsymbol{k})$
and the corresponding eigenvector is given by
\[
\tilde{\mathbf{u}}(\boldsymbol{k})\rightarrow\tilde{\mathbf{u}}(\boldsymbol{k})=\frac{|\tilde{\boldsymbol{B}}(\boldsymbol{k})|}{\sqrt{2}|\tilde{\mathbf{c}}_{\boldsymbol{k}+\boldsymbol{G}(\boldsymbol{k})}|}\left(\begin{array}{c}
\tilde{\mathbf{c}}_{\boldsymbol{k}+\boldsymbol{G}(\boldsymbol{k})}e^{i\boldsymbol{k}\cdot\boldsymbol{\rho}_{1}}\\
\tilde{\mathbf{c}}_{\boldsymbol{k}+\boldsymbol{G}(\boldsymbol{k})}e^{i\boldsymbol{k}\cdot\boldsymbol{\rho}_{2}}
\end{array}\right)\ .
\]
On the other hand, if the folded mode in Eq.~(\ref{eq:Eigenmode_k})
is degenerate, i.e. there are other modes that share its wave vector
and frequency, then it is possible that neither $|\tilde{\mathbf{c}}_{\boldsymbol{k}}|=0$
nor $|\tilde{\mathbf{c}}_{\boldsymbol{k}+\boldsymbol{G}(\boldsymbol{k})}|=0$,
and hence both $\tilde{\mathbf{c}}_{\boldsymbol{k}}$ and $\tilde{\mathbf{c}}_{\boldsymbol{k}+\boldsymbol{G}(\boldsymbol{k})}$
represent correct unfolded eigenmodes, of which we may consider $\tilde{\mathbf{u}}(\boldsymbol{k})$
in Eq.~(\ref{eq:Eigenmode_k}) as a mix. We can ``unmix' ' the degenerate
folded eigenmodes by assigning one unfolded eigenmode to each of the
former. For example, in the special case where $\boldsymbol{Q}_{n}=-\boldsymbol{G}_{\text{tran}}/2$,
the modes at each $\boldsymbol{k}$ are doubly degenerate and can
be represented as $\tilde{\mathbf{u}}_{1}(\boldsymbol{k})$ and $\tilde{\mathbf{u}}_{2}(\boldsymbol{k})$.
In that case, we have
\begin{align*}
\tilde{\mathbf{u}}_{1}(\boldsymbol{k}) & \rightarrow\tilde{\mathbf{u}}_{1}(\boldsymbol{k})=\frac{|\tilde{\boldsymbol{B}}(\boldsymbol{k})|}{\sqrt{2}|\tilde{\mathbf{c}}_{\boldsymbol{k}}|}\left(\begin{array}{c}
\tilde{\mathbf{c}}_{\boldsymbol{k}}e^{i\boldsymbol{k}\cdot\boldsymbol{\rho}_{1}}\\
\tilde{\mathbf{c}}_{\boldsymbol{k}}e^{i\boldsymbol{k}\cdot\boldsymbol{\rho}_{2}}
\end{array}\right)\\
\tilde{\mathbf{u}}_{2}(\boldsymbol{k}) & \rightarrow\tilde{\mathbf{u}}_{2}(\boldsymbol{k})=\frac{|\tilde{\boldsymbol{B}}(\boldsymbol{k})|}{\sqrt{2}|\tilde{\mathbf{c}}_{\boldsymbol{k}+\boldsymbol{G}(\boldsymbol{k})}|}\left(\begin{array}{c}
\tilde{\mathbf{c}}_{\boldsymbol{k}+\boldsymbol{G}(\boldsymbol{k})}e^{i\boldsymbol{k}\cdot\boldsymbol{\rho}_{1}}\\
\tilde{\mathbf{c}}_{\boldsymbol{k}+\boldsymbol{G}(\boldsymbol{k})}e^{i\boldsymbol{k}\cdot\boldsymbol{\rho}_{2}}
\end{array}\right)
\end{align*}
and the unfolded wave vectors of $\tilde{\mathbf{u}}_{1}(\boldsymbol{k})$
and $\tilde{\mathbf{u}}_{2}(\boldsymbol{k})$ are $\boldsymbol{k}$
and $\boldsymbol{k}+\boldsymbol{G}(\boldsymbol{k})$, respectively. 

\begin{figure}
\includegraphics[width=8cm]{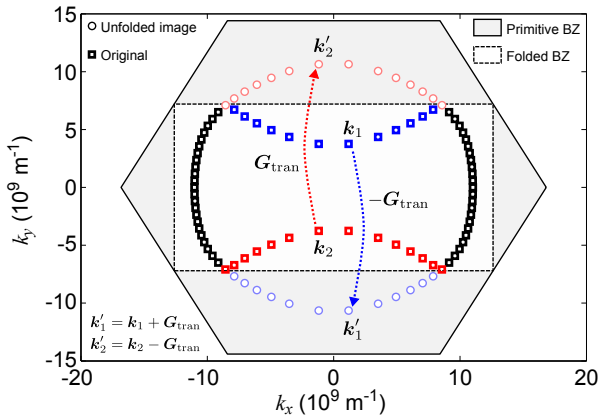}

\caption{Plot of computed ZA phonon modes at $\omega=33$ meV in the folded
Brillouin zone (BZ) and their image points in the unfolded primitive
BZ. The locus of the phonon channels (square symbols) within the folded
BZ forms the shape of a dual-blade ax head while the shape of the
locus of the phonon channels within the primitive BZ is approximately
circular. }
\label{fig:PhononUnfoldingScheme}
\end{figure}

To illustrate the unfolding method, we compute the flexural acoustic
(ZA) phonon channels for $\boldsymbol{Q}_{n}=\frac{2n-N-2}{2N}\boldsymbol{G}_{\text{tran}}$,
where $n=1,\ldots,N$, at $\omega=33$ meV for a bulk graphene system
consisting of $N=24$ 4-atom supercells, like those in Fig.~\ref{fig:SliceSchematics}(a),
in the transverse (armchair) direction. The locus ($\boldsymbol{k}$)
of these phonon channels in the folded BZ is represented by the square
symbols in Fig.~\ref{fig:PhononUnfoldingScheme} and has the shape
of a dual-blade ax head because of the zone-folding of some of the
phonon modes (red and blue square symbols in Fig.~\ref{fig:PhononUnfoldingScheme}).
After applying the Boykin-Klimeck zone-unfolding method,~\citep{TBoykin:PRB05_Practical,TBoykin:PhysicaE09_Non}
the resultant locus of these wave vector points has the approximate
shape of a circle, with the `unfolded' modes represented by circles
in Fig.~\ref{fig:PhononUnfoldingScheme}. The locus of the phonon
channels in which the wave vectors in the folded BZ and their image
in the primitive BZ differ by $\pm\boldsymbol{G}_{\text{tran}}$ is
represented by red ($\boldsymbol{k}^{\prime}=\boldsymbol{k}+\boldsymbol{G}_{\text{tran}}$)
and blue ($\boldsymbol{k}^{\prime}=\boldsymbol{k}-\boldsymbol{G}_{\text{tran}}$)
circles in the primitive BZ and by squares in the folded BZ. For example,
the unfolded points in the primitive BZ at $\boldsymbol{k}_{1}^{\prime}$
and $\boldsymbol{k}_{2}^{\prime}$ in Fig.~\ref{fig:PhononUnfoldingScheme}
are obtained by a displacement of $-\boldsymbol{G}_{\text{tran}}$
and $\boldsymbol{G}_{\text{tran}}$ in reciprocal space, respectively. 

\subsection{Chirality dependence of phonon boundary scattering in graphene}

We study the effects of the edge chirality or orientation on the boundary
scattering of low-energy flexural acoustic (ZA) phonons in graphene.
It is shown by Wei, Chen and Dames in Ref.~\citep{ZWei:JAP12_Wave}
using wave packet dynamics simulations that the scattering of ZA phonons
by the armchair edge can lead to what they call ``wave packet splitting'',
a phenomenon in which the incoming wave packet is split into two or
more outgoing components with dissimilar wave vectors and back-scattered
wave packets are generated after scattering. In the scattering framework,
the two outgoing wave packet components correspond to having two outgoing
phonon channels in which the transition probability is not zero. Wave
packet splitting is however not observed in their simulations of scattering
with the zigzag edge,~\citep{ZWei:JAP12_Wave} suggesting that the
edge chirality exerts a profound effect on the phonon scattering specularity.
Additional evidence of this edge chirality dependence is provided
by molecular dynamics simulations showing that the thermal conductivity
is lower for armchair-edge graphene nanoribbons than for zigzag-edge
graphene nanoribbons.~\citep{JHu:NL09_Thermal} To explain their
findings,~\citep{ZWei:JAP12_Wave} Wei, Chen and Dames attribute
the wave packet splitting to ``the deeper symmetry properties of armchair
and zigzag edges of the hexagonal graphene lattice''.

To understand the physics underlying this phenomenon more precisely,
we investigate the edge scattering of ZA phonons by using our $S$-matrix
approach to compute the transition probabilities between an incoming
ZA phonon channel incident on the edge and the outgoing (reflected)
ZA phonon channels for different edge chirality types. The scope of
our investigation is limited to ZA phonons because the wave packet
splitting of the longitudinal (LA) and transverse acoustic (TA) phonons
can also arise from polarization conversion which does not affect
ZA phonons but can obscure the specularity dependence on edge chirality.
Our simulated system comprises a semi-infinite graphene sheet that
is terminated on the right like in Figs.~\ref{fig:GrapheneEdgeScatteringResults}(a)
and \ref{fig:GrapheneEdgeScatteringResults}(d). In our scattering
calculations, we set $\omega=33$ meV or $5\times10^{13}$ rad/s and
set the incident phonon to be at either normal ($k_{y}=0$) or oblique
($k_{y}\neq0$) incidence. 

\subsubsection{Zigzag edge}

Figure~\ref{fig:GrapheneEdgeScatteringResults}(b) shows the transition
probability distribution along the reciprocal-space locus of the outgoing
ZA phonon channels (solid square symbols) as well as the position
of the incoming phonon channel at $\boldsymbol{k_{1}}$ (solid circle),
which is at normal incidence ($k_{y}=0$) to the zigzag-edge boundary
as shown in Fig.~\ref{fig:GrapheneEdgeScatteringResults}(a). We
find that incident phonon is specularly scattered, i.e. $\mathcal{P}(\boldsymbol{k_{1}})=1$,
to the outgoing phonon channel at $\boldsymbol{\bar{k}_{1}}=\sigma\boldsymbol{k_{1}}$,
where $\sigma$ is the operator corresponding to the reflection $(k_{x},k_{y})\rightarrow(-k_{x},k_{y})$
in reciprocal space, given the computed transition probability of
$P(\boldsymbol{k_{1}}\rightarrow\boldsymbol{\bar{k}_{1}})=1.000$.
Figure~\ref{fig:GrapheneEdgeScatteringResults}(c) shows the transition
probability distribution for the incoming phonon channel at $\boldsymbol{k_{2}}$
which is at an oblique incidence ($k_{y}\neq0$) to the boundary.
The calculation also yields $P(\boldsymbol{k_{2}}\rightarrow\boldsymbol{\bar{k}_{2}})=1.000$
for $\boldsymbol{\bar{k}_{2}}=\sigma\boldsymbol{k_{2}}$, indicating
that the phonon is also specularly scattered. These results are consistent
with the findings in Ref.~\citep{ZWei:JAP12_Wave} where it is shown
that ZA phonon scattering with the zigzag edge is always specular
regardless of the angle of incidence.

\subsubsection{Armchair edge}

\begin{figure*}
\includegraphics[scale=0.4]{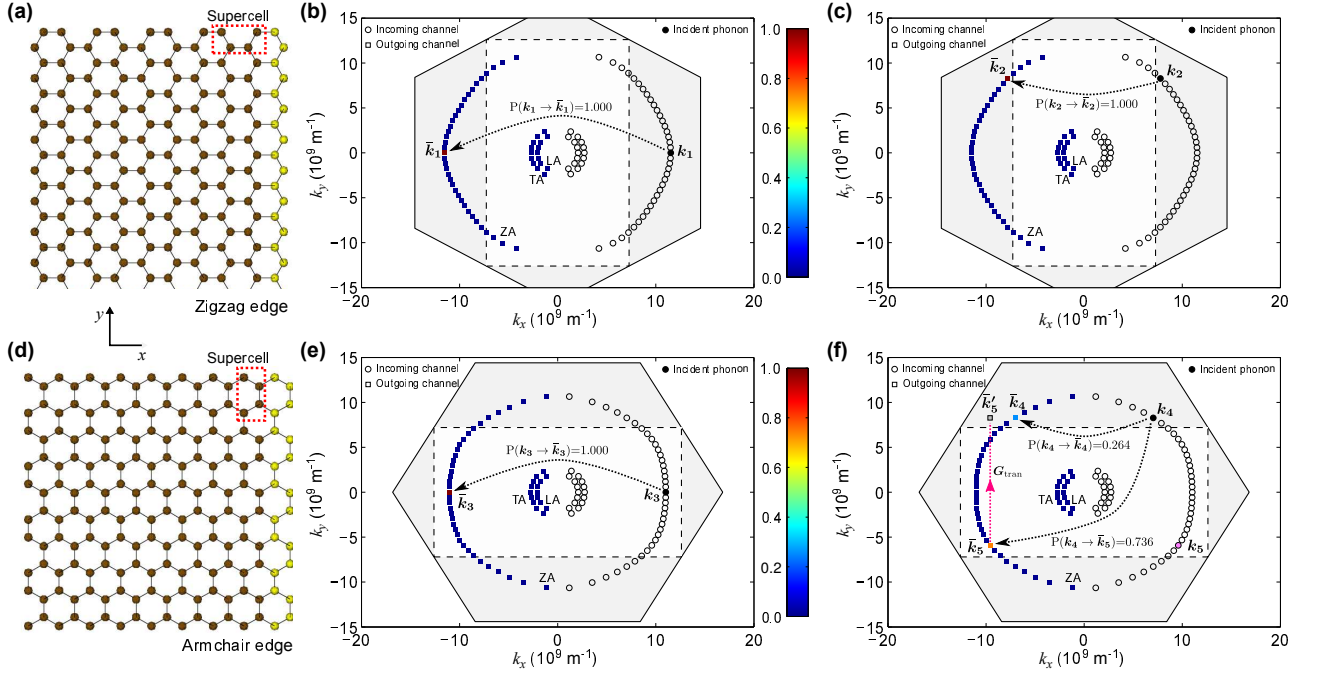}

\caption{Diagram of the smooth graphene \textbf{(a)} zigzag and \textbf{(d)}
armchair edge structure along the $y$ direction. The bulk and edge
C atoms are colored brown and yellow, respectively. For the zigzag
edge, the two-dimensional distribution of the incoming and outgoing
(reflected) phonon channels at $\omega=33$ meV in the first Brillouin
zone (BZ) are shown for an incident flexural phonon at \textbf{(b)}
$\boldsymbol{k_{1}}$ and \textbf{(c)} $\boldsymbol{k_{2}}$. The
locus of channels can be placed in three `rings' corresponding to
the longitudinal acoustic (LA), transverse acoustic (TA) and flexural
acoustic (ZA) phonons. In addition to the first BZ, we also draw the
outline of the folded BZ (dashed lines) corresponding to the supercell
shown in \textbf{(a)}. The associated transition probabilities for
the outgoing reflected channels are indicated in color. The incident
phonons are specularly scattered to the channels at $\boldsymbol{\bar{k}_{1}}$
and $\boldsymbol{\bar{k}_{2}}$. For the armchair edge, the distributions
of the incoming and outgoing phonon channels are similarly shown for
the incident phonon at \textbf{(e)} $\boldsymbol{k_{3}}$ and \textbf{(f)}
$\boldsymbol{k_{4}}$. The phonon at $\boldsymbol{k_{3}}$ is specularly
scattered to $\boldsymbol{\bar{k}_{3}}$ but the phonon at $\boldsymbol{k_{4}}$
is scattered into two channels at $\boldsymbol{\bar{k}_{4}}$ and
$\boldsymbol{\bar{k}_{5}}$, with the $\boldsymbol{k_{4}}\rightarrow\boldsymbol{\bar{k}_{5}}$
transition probability \emph{greater} than the $\boldsymbol{k_{4}}\rightarrow\boldsymbol{\bar{k}_{4}}$
transition probability. The point at $\boldsymbol{\bar{k}_{5}^{\prime}}=\boldsymbol{\bar{k}_{5}}+\boldsymbol{G}_{\text{tran}}$
is collinear with $\boldsymbol{k_{4}}$ and $\boldsymbol{\bar{k}_{4}}$
as a consequence of Eq.~(\ref{eq:EdgeMomentumConservation}). }

\label{fig:GrapheneEdgeScatteringResults}
\end{figure*}

We repeat our calculations for ZA phonon scattering with the armchair
edge as shown in Fig.~\ref{fig:GrapheneEdgeScatteringResults}(d).
At normal incidence to the armchair edge, the incident phonon at $\boldsymbol{k_{3}}$
is specularly scattered to the outgoing phonon channel $\boldsymbol{\bar{k}_{3}}=\sigma\boldsymbol{k_{3}}$
since $P(\boldsymbol{k_{3}}\rightarrow\boldsymbol{\bar{k}_{3}})=1.000$
as shown in Fig.~\ref{fig:GrapheneEdgeScatteringResults}(e). However,
at oblique incidence, the scattering of the incoming phonon channel
at $\boldsymbol{k_{4}}$ is only partially specular as $P(\boldsymbol{k_{4}}\rightarrow\boldsymbol{\bar{k}_{4}})=0.264$
for $\boldsymbol{\bar{k}_{4}}=\sigma\boldsymbol{k_{4}}$ and the incident
phonon is also backscattered to a second outgoing phonon channel at
$\boldsymbol{\bar{k}_{5}}$ with $P(\boldsymbol{k_{4}}\rightarrow\boldsymbol{\bar{k}_{5}})=0.736$.
There are no other outgoing channels to which the incident phonon
is scattered because the total transition probability of these two
outgoing channels is $P(\boldsymbol{k_{4}}\rightarrow\boldsymbol{\bar{k}_{4}})+P(\boldsymbol{k_{4}}\rightarrow\boldsymbol{\bar{k}_{5}})=1.000$.
This splitting of the incident ZA phonon to two outgoing ZA phonon
channels after scattering with the armchair edge is qualitatively
consistent with the wave packet splitting observed in Ref.~\citep{ZWei:JAP12_Wave}. 

To explain the partial scattering specularity of the incident phonon
at $\boldsymbol{k_{4}}$, we note that the $y$ component of $\boldsymbol{\bar{k}_{5}}-\boldsymbol{\bar{k}_{4}}$,
which is the difference in the reciprocal-space position of the outgoing
phonon channels at $\boldsymbol{\bar{k}_{4}}$ and $\boldsymbol{\bar{k}_{5}}$,
is equal to $\boldsymbol{G}_{\text{tran}}$ which characterizes the
periodicity of the armchair edge as well as that of the supercell
{[}Fig.~\ref{fig:GrapheneEdgeScatteringResults}(d){]} in the transverse
($y$) direction. To make this clearer, we plot in Fig.~\ref{fig:GrapheneEdgeScatteringResults}(f)
the point $\boldsymbol{\bar{k}_{5}^{\prime}}=\boldsymbol{\bar{k}_{5}}+\boldsymbol{G}_{\text{tran}}$
which is collinear with $\boldsymbol{k_{4}}$ and $\boldsymbol{\bar{k}_{4}}$.
More generally, this implies that any elastic phonon scattering by
the edge must satisfy the conservation condition
\begin{equation}
\hat{\boldsymbol{y}}\cdot(\boldsymbol{k}_{\text{in}}-\boldsymbol{k}_{\text{out}})=m|\boldsymbol{G}_{\text{tran}}|\label{eq:EdgeMomentumConservation}
\end{equation}
where $m\in\mathbb{Z}$ and $\boldsymbol{k}_{\text{in}}$ ($\boldsymbol{k}_{\text{out}}$)
is the wave vector of the incoming (outgoing) phonon channel. 

Therefore, given Eq.~(\ref{eq:EdgeMomentumConservation}), we can
explain why phonon scattering by the armchair edge is fully specular
in Fig.~\ref{fig:GrapheneEdgeScatteringResults}(e) and only partially
specular in Fig.~\ref{fig:GrapheneEdgeScatteringResults}(f). In
Fig.~\ref{fig:GrapheneEdgeScatteringResults}(e) where the incoming
phonon at $\boldsymbol{k_{3}}$ is at normal incidence to the boundary,
the only outgoing phonon channel that satisfies Eq.~(\ref{eq:EdgeMomentumConservation})
is at $\boldsymbol{\bar{k}_{3}}$ and hence, the incident phonon undergoes
fully specular scattering. On the other hand, when the incoming phonon
is at $\boldsymbol{k_{4}}$, there are two outgoing phonon channels
($\boldsymbol{\bar{k}_{4}}$ and $\boldsymbol{\bar{k}_{5}}$) that
satisfy Eq.~(\ref{eq:EdgeMomentumConservation}), such that $\hat{\boldsymbol{y}}\cdot(\boldsymbol{k_{4}}-\boldsymbol{\bar{k}_{4}})=0$
and $\hat{\boldsymbol{y}}\cdot(\boldsymbol{k_{4}}-\boldsymbol{\bar{k}_{5}})=|\boldsymbol{G}_{\text{tran}}|$,
resulting in a ``splitting'' of the incoming phonon. 

Along the same lines, we can also explain the full specularity of
ZA phonon scattering and the absence of wave packet splitting for
the zigzag edge. The greater symmetry of the zigzag edge means that
its $|\boldsymbol{G}_{\text{tran}}|$ is larger than the $|\boldsymbol{G}_{\text{tran}}|$
of the armchair edge since $|\boldsymbol{G}_{\text{tran}}|=\frac{2\pi}{\sqrt{3}a}$
and $\frac{2\pi}{3a}$ for the zigzag and armchair edge, respectively,
where $a$ is the carbon-carbon bond length. This can also be seen
when we compare the width of the folded BZ along the $k_{y}$-axis
in Figs.~\ref{fig:GrapheneEdgeScatteringResults}(b) and \ref{fig:GrapheneEdgeScatteringResults}(e).
Hence, the conservation condition in Eq.~(\ref{eq:EdgeMomentumConservation})
is more restrictive for the zigzag edge because its larger $|\boldsymbol{G}_{\text{tran}}|$
allows for only one outgoing phonon channel when $\omega=33$ meV. 

\subsection{Effect of graphene edge chirality and isotopic disorder on ZA phonon
specularity }

\subsubsection{Ordered edges}

We use our $S$-matrix method to study how the ZA phonon boundary
scattering specularity ($\mathcal{P}$) varies systematically with
frequency ($\omega$) and wave vector ($\boldsymbol{k}$) for different
edge chirality configurations. The specularity parameter distribution
of the incoming flexural acoustic (ZA) phonons is computed at $\omega=l\omega_{0}$,
where $\omega_{0}=6.6$ meV or $10^{13}$ rad/s and $l=1,\ldots,6$,
in Fig.~\ref{fig:ZAPhononSpecularity} for: (a) the ideal zigzag
edge with $N=42$ unit cells or 84 atoms and (b) the ideal armchair
edge with $N=24$ unit cells or 96 atoms in the transverse direction.
At each frequency point, the locus of all the incoming ZA phonons
is represented by a constant-frequency arc, as shown in Fig.~\ref{fig:ZAPhononSpecularity},
and the loci form a concentric arrangement of arcs with the innermost
and outermost arc corresponding to $\omega=\omega_{0}$ and $\omega=6\omega_{0}$,
respectively. 

Figure~\ref{fig:ZAPhononSpecularity}(a), which corresponds to the
ideal zigzag edge, shows that the specularity is perfect ($\mathcal{P}=1$)
as expected for all incoming ZA phonons in the frequency range studied,
confirming the conservation condition in Eq.~(\ref{eq:EdgeMomentumConservation}).
However, in Fig.~\ref{fig:ZAPhononSpecularity}(b) which corresponds
to the ideal armchair edge, the the ZA phonon specularity varies with
the frequency $\omega$ and wave vector $\boldsymbol{k}=(k_{x},k_{y})$,
in agreement with the findings of Ref.~\citep{ZWei:JAP12_Wave}.
Figure~\ref{fig:ZAPhononSpecularity}(b) shows that the variation
in specularity with $\boldsymbol{k}$ becomes more pronounced at larger
$\omega$. In each constant-frequency arc in Fig.~\ref{fig:ZAPhononSpecularity}(b)
for $\omega=4\omega_{0}$ to $6\omega_{0}$, $\mathcal{P}(\boldsymbol{k})$
approaches its \emph{minimum} as $k_{y}$ approaches $\pm\frac{1}{2}|\boldsymbol{G}_{\text{tran}}|$
as indicated in Fig.~\ref{fig:ZAPhononSpecularity}(b). The existence
of this minimum at a particular incident angle is reported but not
explained in Ref.~\citep{ZWei:JAP12_Wave}. 

For the specularity minimum at $k_{y}=\pm\frac{1}{2}|\boldsymbol{G}_{\text{tran}}|$,
there are two outgoing channels at $\boldsymbol{\bar{k}}$ and $-\boldsymbol{k}$.
Figure~\ref{fig:ZAPhononSpecularity}(b) shows that as we increase
the frequency, the $\boldsymbol{k}\rightarrow-\boldsymbol{k}$ transition,
which corresponds to the reversal of the phonon trajectory such that
the angle of incidence is equal to the \emph{negative} of the angle
of reflection, becomes increasingly more probable. This implies that
at high phonon frequencies, the a greater proportion of the phonon
momentum in the $y$-direction is lost due to scattering with the
\emph{ideal} armchair edge.

\subsubsection{Disordered edges}

\begin{figure*}
\includegraphics[scale=0.4]{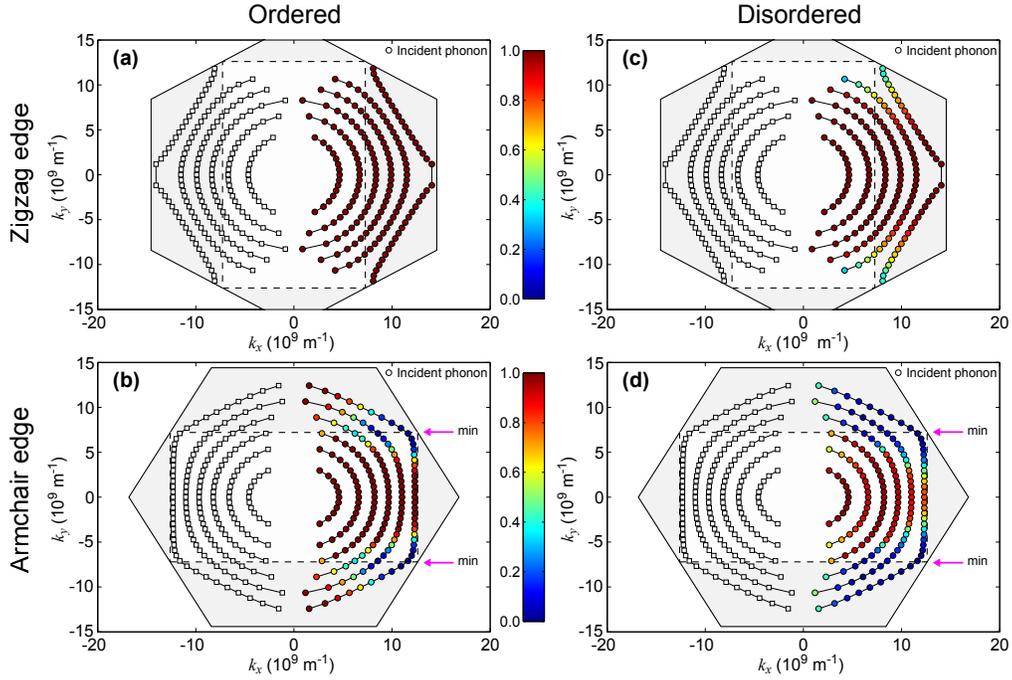}

\caption{Plot of the specularity parameter value $\mathcal{P}$ distribution
within the Brillouin zone for incoming (incident) flexural acoustic
(ZA) phonon channels (filled circles) at $\omega=l\omega_{0}$, where
$\omega_{0}=6.6$ meV or $10^{13}\ \text{rad s}^{-1}$ and $l=1,\ldots,6$,
for the \textbf{(a)} ordered zigzag, \textbf{(b)} ordered armchair,
\textbf{(c)} disordered zigzag and \textbf{(d)} disordered armchair
edge, computed using Eq.~(\ref{eq:LeftLeadSpecularityParameterDefn}).
The outgoing ZA phonon channels are represented by hollow square symbols.
The ZA phonon channels are arranged in concentric constant-$\omega$
arcs. The $\mathcal{P}(\boldsymbol{k})$ minima at each $\omega$
lie along the $k_{y}=\pm\frac{1}{2}|\boldsymbol{G}_{\text{tran}}|$
line (labeled `min') for both the isotopically ordered and disordered
armchair edge.}

\label{fig:ZAPhononSpecularity}
\end{figure*}

Given the role of the edge translational symmetry in the ZA phonon
scattering specularity, it would be interesting to see the effect
of the loss of that symmetry on phonon specularity. To break the translational
symmetry of the graphene edge, we randomly replace 25 percent of the
edge$^{12}$C atoms with $^{24}$C atoms {[}Figs.~\ref{fig:ZAPhononSpecularity}(a)
and \ref{fig:ZAPhononSpecularity}(d){]} to create isotopic disorder
along the edges. 

Figure~\ref{fig:ZAPhononSpecularity} shows the specularity parameter
distribution at $\omega=l\omega_{0}$, where $l=1,\ldots,6$, for
incoming ZA phonon channels at: (a,b) the zigzag edge with $N=42$
unit cells or 84 atoms and (c,d) the armchair edge with $N=24$ unit
cells or 96 atoms in the transverse direction. The specularity distributions
for the mass-disordered edges in Fig.~\ref{fig:ZAPhononSpecularity}(b)
and (d) are obtained after averaging over 20 realizations of disorder
while the distributions in Figs.~\ref{fig:ZAPhononSpecularity}(a)
and \ref{fig:ZAPhononSpecularity}(c) have no disorder and represent
the baseline specularity values. 

A comparison of Figs.~\ref{fig:ZAPhononSpecularity}(a) and \ref{fig:ZAPhononSpecularity}(c)
shows that the $\mathcal{P}(\boldsymbol{k})=1$ result no longer holds
in the disordered zigzag edge. We observe that the specularity decreases
as the frequency and the angle of incidence decrease. This dependence
on the angle of incidence is unexpected as models of surface roughness
scattering~\citep{JZiman:Book60_Electrons,AMaznev:PRB15_Boundary}
suggest that the specularity should decrease monotonically with the
angle of incidence. This suggests that the effect of edge disorder
is different from that of edge roughness and that caution should be
exercised when using specularity approximations based on surface roughness
scattering. 

In Fig.~\ref{fig:ZAPhononSpecularity}(d) at large $\omega$ ($\omega=l\omega_{0}$
for $l=4$ to $6$), we observe that the specularity parameter $\mathcal{P}(\boldsymbol{k})$
is maximum at normal incidence to the edge but decreases as the angle
of incidence increases before reaching its minimum when $k_{y}=\pm\frac{1}{2}|\boldsymbol{G}_{\text{tran}}|$
like in Fig.~\ref{fig:ZAPhononSpecularity}(c). Comparing Figs.~\ref{fig:ZAPhononSpecularity}(c)
and (d), we find that the isotopic disorder at the armchair edge reduces
$\mathcal{P}(\boldsymbol{k})$, with the decrease in $\mathcal{P}(\boldsymbol{k})$
becoming larger at higher frequency and angle of incidence, similar
to the trend observed for the zigzag edge. 

\section{Summary and conclusion}

We have described the improvement of the atomistic Green's function
(AGF) method for treating individual phonon transmission and reflection,
and shown explicitly how the phonon transmission and reflection matrices
can be determined numerically and used to construct the unitary $S$
matrix that characterizes scattering by the interface and treats bulk
phonon modes as scattering channels. In our AGF-based $S$-matrix
approach, the scattering amplitude between the phonon channels is
determined from the corresponding $S$-matrix element and yields the
transition probability for the forward (transmission) or backward
(reflection) scattering process. We illustrate the advantages of our
new approach by first applying it to the example of phonon scattering
at the junction of two isotopically different (8,8) carbon nanotubes.
The $S$-matrix approach allows us to determine the dependence of
the phonon transmission and reflection on frequency, polarization
and phonon velocity. We also analyze the transition probability for
individual scattering processes as well as describe the role of intra
and inter-subband processes in phonon reflection. 

We also illustrate the utility of the method by applying it to the
study of phonon reflection from a graphene edge. We take advantage
of the transverse periodic boundary condition to partition the system
into its Fourier components for more efficient computation of matrix
variables such as the surface Green's function. For clarity, the scattering
channels are mapped to the bulk phonon modes of graphene using the
Boykin-Klimeck zone-unfolding technique. Our numerical calculations
reveal that unlike the zigzag edge, phonon scattering with the armchair
edge is only partially specular because of the symmetry difference
between the armchair edge and the bulk lattice. We also find that
the specularity varies with wave vector and frequency and decreases
as expected when isotopic disorder is introduced to the edge. 

Potentially, the application of our AGF-based $S$-matrix method in
the atomistic simulations of other interfaces can provide a similarly
detailed picture of phonon transmission and reflection, and shed light
on the relationship between phonon scattering and the atomistic structure
of the interface or surface. The method may also be incorporated into
multiscale models of phonon and thermal conduction in heterogeneous
solids with interfaces~\citep{DSingh:JHT11_Effect} by combining
it with the transport models based on the Boltzmann transport equation.
The method can also be used to estimate phonon specularity in transport
models of low-dimensional systems (e.g. silicon nanowires or graphene
nanoribbons) in which edge scattering is important for momentum relaxation.
In addition, the formalism presented in this paper may be applicable
on its own to the numerical simulation of scattering in linear systems
(e.g. photonic crystals~\citep{JJoannopoulos:Book08_Photonic}) that
have a lattice structure and are second order in time. 
\begin{acknowledgments}
This work was supported in part by a grant from the Science and Engineering
Research Council (Grant No. 152-70-00017) and financial support from
the Agency for Science, Technology and Research (A{*}STAR), Singapore.
I also gratefully acknowledge the gracious hospitality shown by the
Department of Materials Science and Metallurgy at the University of
Cambridge where part of this work was carried out. 
\end{acknowledgments}

\bibliographystyle{apsrev4-1}
\bibliography{PhononNotes}

\end{document}